\documentclass[12pt,twoside]{article}

\usepackage{nicefrac}

\usepackage[ansinew]{inputenc}
\usepackage{amssymb,latexsym,amsfonts,url,graphicx,amsmath}
\usepackage[english]{babel}

\usepackage{apacite}

\newcommand{\mb}{\mathbf}
\newcommand{\av}{{\mathbf{a}}}
\newcommand{\Av}{{\mathbf{A}}}
\newcommand{\Dv}{{\mathbf{D}}}
\newcommand{\zerov}{{\mathbf{0}}}
\newcommand{\mbI}{{\mathbf{I}}}
\newcommand{\Xv}{{\mathbf{X}}}
\newcommand{\yv}{\mathbf{y}}
\newcommand{\xv}{{\mathbf{x}}}
\newcommand{\Cv}{{\mathbf{C}}}
\newcommand{\verror}{\sigma^2}

\newcommand{\psiv}{{\boldsymbol{\psi}}}
\newcommand{\alphav}{{\boldsymbol{\alpha}}}
\newcommand{\deltav}{{\boldsymbol{\delta}}}
\newcommand{\varthetav}{{\boldsymbol{\vartheta}}}
\newcommand{\Sigmav}{\boldsymbol{\Sigma}}
\DeclareMathOperator{\diag}{diag} 

\DeclareMathOperator{\Var}{var}
\newcommand{\Gammainv}[1]{\mathcal{G}^{-1} \left(#1\right)}
\newcommand{\Normal}[1]{\mathcal{N}\left(#1\right)}

\newcommand{\Normalpdfa}[2]{\pdf{N}{#1}{#2}}

\newcommand{\pdf}[3]{f_{   #1 }(#2;#3)}
\newcommand{\Betadis}[1]{\mathcal{B}\left(#1\right)}
\newcommand{\slab}{\text{slab}}
\newcommand{\spike}{\text{spike}}

\begin{document}
\renewcommand{\BBAA}{and}
\thispagestyle{empty}

~

\vspace{-26mm} \noindent

\vspace{2mm}
\begin{center}
{\Large\bf Comparing Spike and Slab Priors for \\[1mm]
           Bayesian Variable Selection}\\[5mm]
{\large Gertraud Malsiner-Walli and Helga Wagner \\[2mm]
        Johannes Kepler Universität Linz, Austria}
\end{center}

\begin{quote}
{\bf Abstract:} {An important task in building regression models
is to decide which regressors should be included in the final
model. In a Bayesian approach, variable selection can be performed
using mixture priors with a spike and a slab component for the
effects subject to selection. As the spike is concentrated at
zero, variable selection is based on the probability of assigning
the corresponding regression effect to the slab component. These
posterior inclusion probabilities can be determined by MCMC
sampling. In this paper we compare the MCMC implementations for
several spike and slab priors with regard to posterior inclusion
probabilities and their sampling efficiency for simulated data.
Further, we investigate posterior inclusion probabilities
analytically for different slabs in two simple settings.
Application of variable selection with spike and slab priors is
illustrated on a data set of psychiatric patients where the goal
is to identify covariates affecting metabolism.}


\end{quote}

\section{Introduction}

A major task in building a regression model is to select  those regressors  from a large set of potential covariates which should be included in the final model. Correct classification of regressors as having (nearly) zero or non-zero effects is
important: omitting regressors with non-zero effect will lead to biased estimates whereas inclusion of regressors with zero effect causes loss in estimation precision and predictive performance of the model.

For the regression coefficients, many Bayesian variable selection methods use mixture priors with two components: a spike concentrated around zero and a comparably flat slab. In this paper we compare spike and slab priors with two different specifications for the spike: absolutely continuous and  spikes defined by a point mass at zero, so called Dirac spikes. We consider here Dirac spikes combined with different normal slabs and priors where both spike and slab are normal distributions as in \citeA{geo-mcc:var} or scale mixtures of normals as in \citeA{ish-rao:spi} and
\citeA{kon-etal:bay}.

Bayesian variable selection with spike and slab priors can be accomplished  by MCMC methods, but depending on the type of the spike  the specific implementations differ: A Dirac spike requires computation of  marginal likelihoods, i.e.\ integrating over the parameters  subject to selection, in each MCMC iteration. This is not necessary for spikes specified by an absolutely continuous distribution. However, regression effects are not shrunk exactly to zero and therefore the dimension of the model is not reduced during MCMC. In this paper we compare posterior inclusion probabilities under different spike and slab priors as well as their MCMC sampling efficiency.

The rest of the paper is structured as follows. Section \ref{sec:model} describes the basic model and the two types of spike and slab priors. Implementation of MCMC sampling schemes is outlined  for both spike types in Section \ref{sec:MCMC} and
Section \ref{sec:Simu} presents results from a simulation study comparing five  different spike and slab priors on simulated data. To get further insight, posterior inclusion probabilities are investigated analytically in two simple settings for Dirac spikes combined with different slabs in Section \ref{sec:theo}. Section \ref{sec:App} illustrates application of Bayesian variable selection on a data set where the goal is to identify covariates which have an effect on metabolism of psychiatric patients. Finally, Section \ref{sec:Sum} summarizes the results and
indicates  modifications for the slab component to be considered in further research.

\section{Model Specification} \label{sec:model}

\subsection{The Linear Regression Model}

In the standard linear regression model the outcome $\yv = (y_1, \dots, y_N)$ of
subjects  $i = 1, \dots, N$ is modeled as a linear function  of the regressors with a Gaussian error term,
\begin{equation}\label{regmod}
 \yv = \mb{1}\mu + \Xv\alphav + \boldsymbol{\varepsilon}\,,
 \qquad \boldsymbol{\varepsilon} \sim \Normal{\zerov, \mbI\verror}\,.
\end{equation}
Here $\alphav$ is the $d \times 1$ vector of regression coefficients. We assume that the covariate vectors  are centered with the null vector as mean, so that $\Xv'\mb{1} = \zerov$ and the mean $\mu$ is constant over all models. As the columns of the design matrix are orthogonal to the unit vector, the log-likelihood can be written as
$$
 l(\yv|\mu, \alphav, \verror)
 = -\frac N2\log(2\pi\verror) - \frac 1{2\verror}
   \Big(N(\bar y - \mu)^2 + (\yv_c - \Xv\alphav)'(\yv_c - \Xv\alphav)\Big)\,,
$$
where $\yv_c = \yv - \mb{1} \bar y $ denotes the vector of centered responses.

In a Bayesian approach, model specification is completed with priors for the model parameters $(\mu, \verror, \alphav)$. We assume a prior of the structure $p(\mu, \verror, \alphav) = p(\mu, \verror) p(\alphav|\verror, \mu)$ with the usual uninformative prior for mean and error variance
\begin{equation}\label{pri}
 p(\mu, \verror) = \frac 1{\verror},
\end{equation}
and use spike and slab priors  for the regression coefficients $\alphav$.

\subsection{Spike and Slab Priors}

Mixture priors with spike and slab components have been used extensively for variable selection, see e.g.~\citeA{mit-bea:bay}, \citeA{geo-mcc:var, geo-mcc:app} and \citeA{ish-rao:spi}. The spike component, which concentrates its mass at values close to zero, allows shrinkage of small effects to zero, whereas the slab component has its mass spread over a wide range of plausible values for the regression coefficients. To specify spike and slab priors we introduce indicator variables $\deltav = (\delta_1, \dots, \delta_d)$ where $\delta_j$ takes the value 1, if $\alpha_j$ is allocated to the slab component and we denote by $\alphav_\deltav$ the vector comprising those elements of $\alphav$ where $\delta_j = 1$. We consider priors, where regression effects allocated to the spike component are independent of each other and independent of $\alphav_\deltav$ a priori, whereas elements of $\alphav_\deltav$ may be dependent. These spike and slab priors can be written as
$$
 p(\alphav|\deltav)
 =
 p_{\slab}(\alphav_\deltav)\prod_{j:\delta_j=0}p_{\spike}(\alpha_j)\,,
$$
where $p_{\spike}$ and $p_{\slab}$ denote the univariate spike and the multivariate slab distribution respectively. The prior inclusion probability $p(\delta_j=1)$ of the effect $\alpha_j$ is specified hierarchically as
$$
 p(\delta_j=1|\omega) = \omega\,,
 \qquad
 \omega \sim \mathcal{B}(a_{\omega}, b_{\omega})\,.
$$
Note, that the indicator variables $\delta_j$ are independent conditional on the prior inclusion probability $\omega$, but dependent marginally. This might not be justified in practical applications and could be relaxed by using an individual inclusion probability $\omega_j$ for each regression effect $\alpha_j$,
$$
 p(\delta_j=1|\omega_j) = \omega_j\,,
 \qquad
 \omega_j \sim \mathcal{B}(a_{\omega_j}, b_{\omega_j})\,.
$$
Prior information on individual inclusion probabilities could be incorporated by appropriate choice of the parameters $a_{\omega_j}$ and $b_{\omega_j}$.

The introduction of indicator variables allows classification of regression effects as (practically) zero, if $\delta_j = 0$ and non-zero otherwise. Variable selection is based on the posterior probability of  assigning the corresponding regression effect to the slab component, i.e.\ the posterior inclusion probability $p(\delta_j=1|\mathbf{y})$, which can be sampled by MCMC methods. Basically two different types of spikes have been proposed in the literature: Spikes specified by an absolutely continuous distribution and spikes specified  by a point mass at zero, called Dirac spikes. Specifications of priors for  both spike types, which are compared in this paper, are presented in more detail in the following sections.

\subsubsection{Absolutely Continuous Spikes}\label{sec:pricon}

To specify an absolutely continuous spike, in principle any unimodal continuous distribution with mode at zero could be used. Usually absolutely continuous spikes are combined with slabs, where the components of $\alphav_\deltav$ are independent
conditional on $\deltav$, i.e.
$$
 p_{\slab}(\alphav_\deltav) = \prod_{j:\delta_j=1} p_{\slab}(\alpha_j)\,.
$$
Here we consider priors where spike and slab components are specified by the same distribution family but with a variance ratio $r$ considerably smaller than 1,
\begin{equation} \label{varrat}
 r = \frac{\Var_{\spike}(\alpha_j)}{\Var_{\slab}(\alpha_j)} <\!\!< 1\,.
\end{equation}
We use only spikes and slabs  which can be represented as scale mixtures of normal distributions with zero mean,
$$
 \alpha_j|\delta_j, \psi_j \sim \Normal{0, r(\delta_j)\psi_j}\,,
 \qquad
 \psi_j|\varthetav \sim p(\psi_j|\varthetav)\,,
$$
where
$$
 r(\delta_j) =
 \begin{cases} r \quad \text{if } \delta_j=0\\
               1 \quad \text{if } \delta_j=1
 \end{cases}
$$
and the distribution of $\psi_j$ may depend on a further parameter $\varthetav$. In particular, we consider normal spikes and slabs with constant $\psi_j \equiv V$ (called SSVS prior) and normal mixtures of inverse Gamma distributions (NMIG prior), where $\psi_j \sim \Gammainv{\nu,Q}$. Priors with normal spikes and slabs were introduced in \citeA{geo-mcc:var} to perform stochastic search variable selection and NMIG spikes and slabs were proposed in \citeA{ish-rao:det} and \citeA{ish-rao:spi} for variable selection in Gaussian regression models and used in \citeA{kon-etal:bay} for survival data. Note that  for the NMIG prior marginally both spike and slab component are student distributions,
$$
 p_{\spike}(\alpha_j) = t_{2\nu}(0, r Q/\nu)
 \qquad\text{and}\qquad
 p_{\slab}(\alpha_j) = t_{2\nu}(0, Q/\nu)\,.
$$

\subsubsection{Dirac Spike}\label{sec:pridir}

A Dirac spike is specified as $p_{\spike}(\alpha_j) = p(\alpha_j|\delta_j=0) = \Delta_0(\alpha_j)$. We combine Dirac spikes with slab components of the form
$$
 p_\slab(\alphav_\deltav)
 = \Normalpdfa{\alphav_\deltav}{\av_{0,\deltav}, \Av_{0,\deltav}\verror}\,,
$$
where $\Normalpdfa{\xv}{\boldsymbol{\mu}, \Sigmav}$ denotes the density of the multivariate $\Normal{\boldsymbol{\mu}, \Sigmav}$-distribution. In particular we use
\begin{itemize}\vspace*{-2mm}
 \item the independence slab (i-slab), where $\av_{0,\deltav} =
       \zerov$ and $\Av_{0,\deltav} = c\mbI$,\vspace*{-2mm}
 \item the g-slab, where $\av_{0,\deltav} = \zerov$ and $\Av_{0,
       \deltav} = g(\Xv_\deltav'\Xv_\deltav)^{-1}$,\vspace*{-2mm}
 \item the fractional slab (f-slab), where $\av_{0,\deltav} =
       (\Xv_\deltav'\Xv_\deltav)^{-1}\Xv_\deltav'\yv_c$ and
       $\Av_{0,\deltav}=1/b\;(\Xv_\deltav'\Xv_\deltav)^{-1}$.
\end{itemize}

\begin{figure}
 \includegraphics[width=5cm]{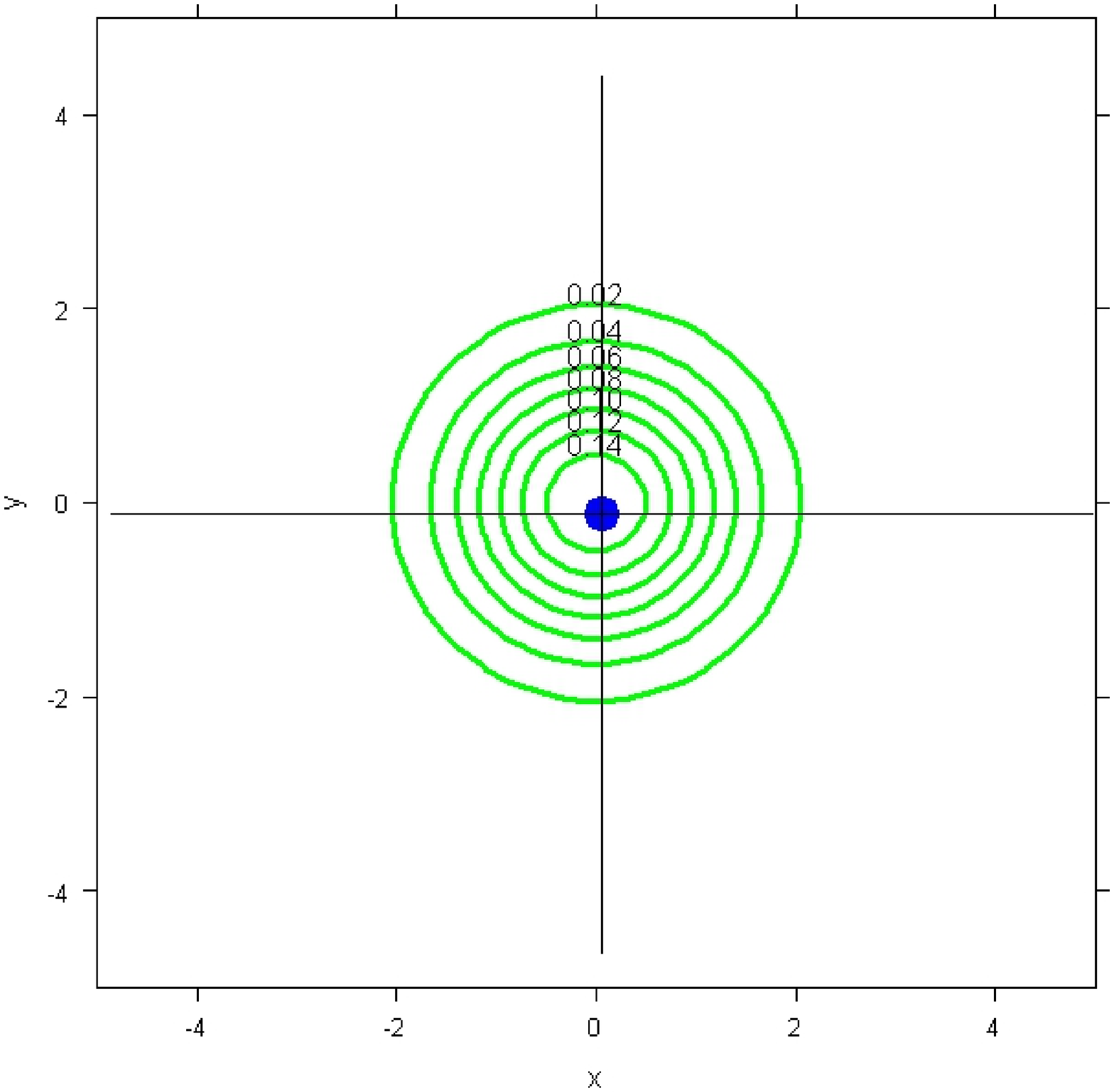}\hspace*{-2mm}
 \includegraphics[width=5cm]{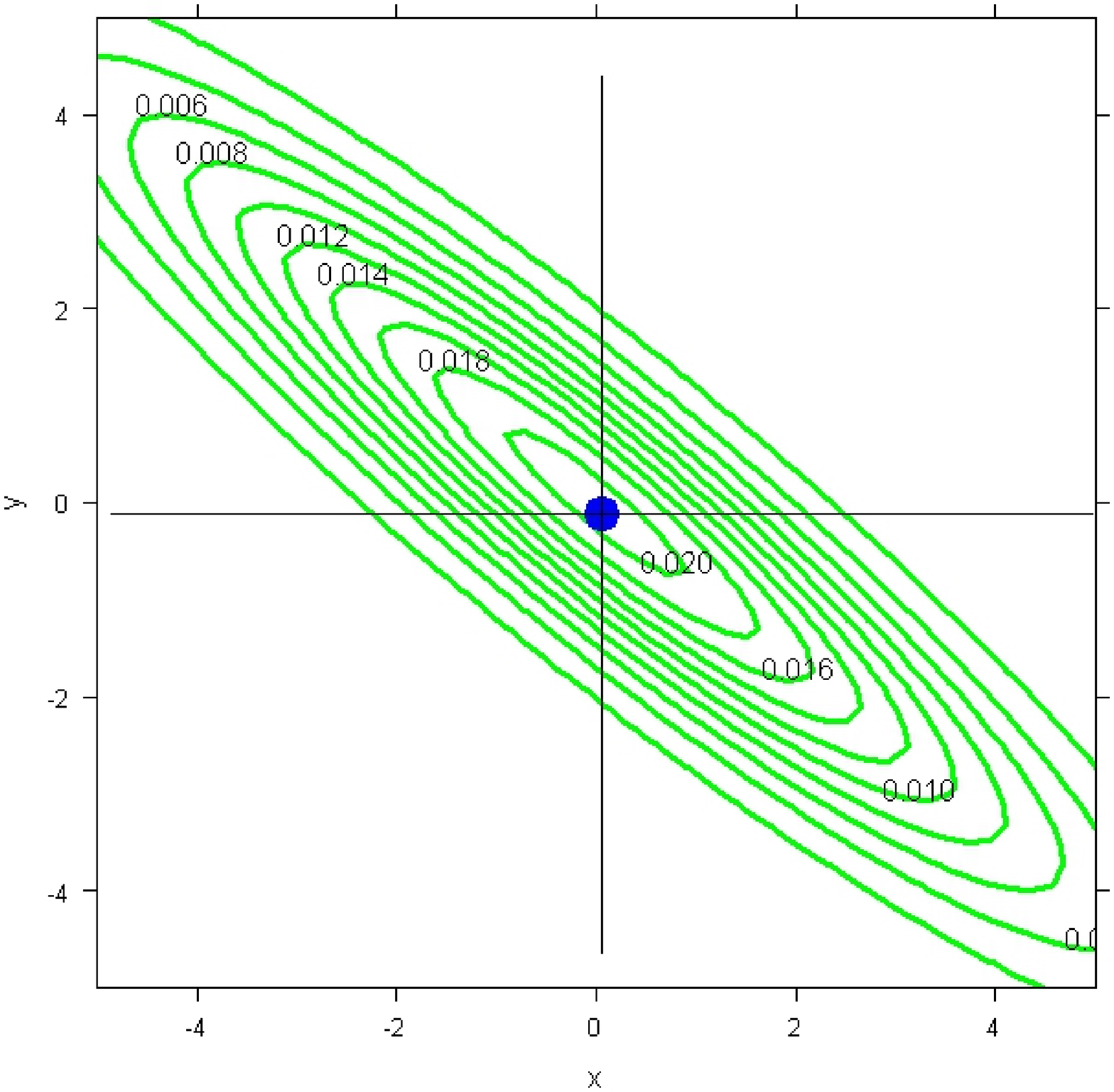}\hspace*{-2mm}
 \includegraphics[width=5cm]{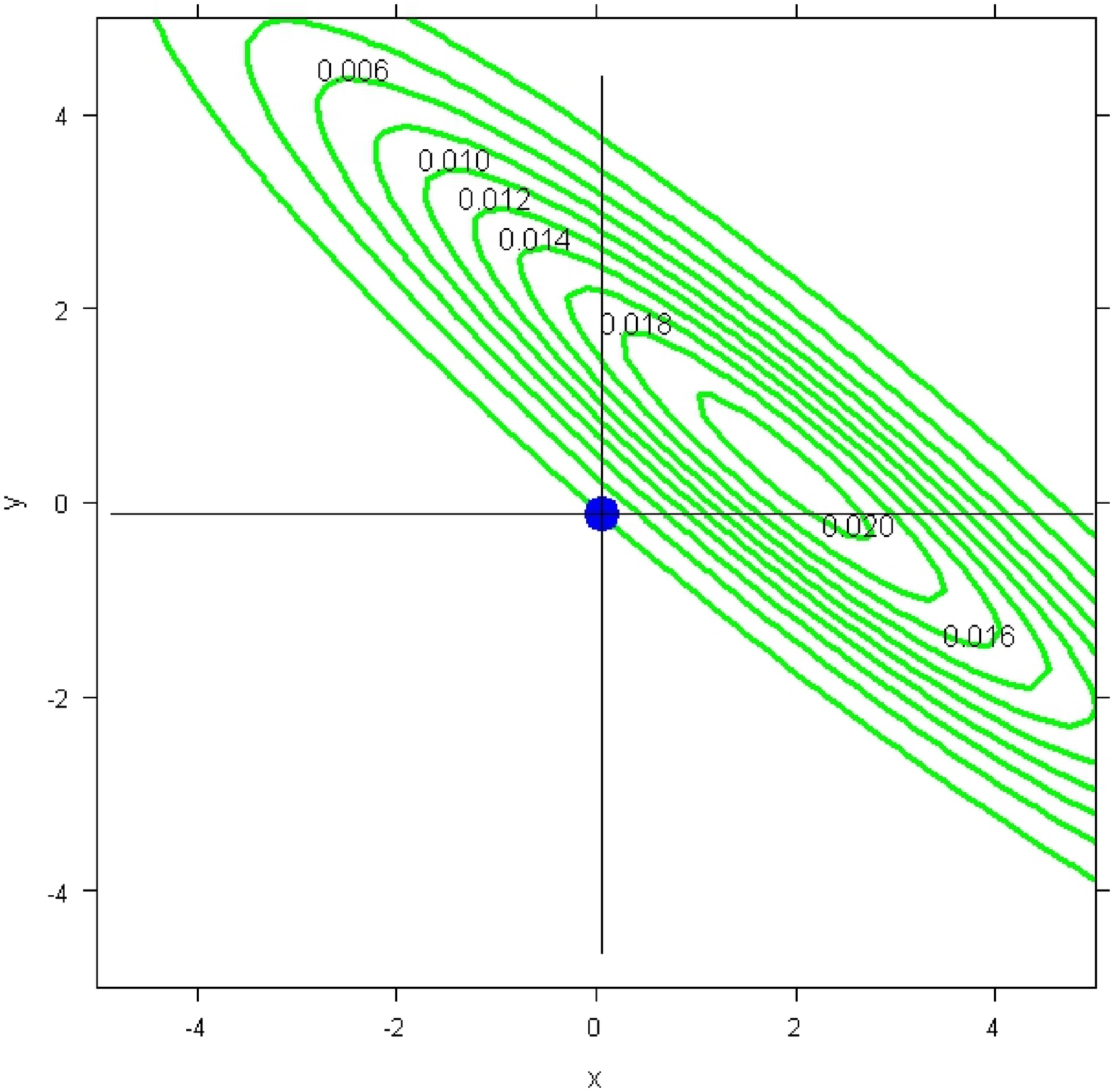}
 \caption{Contour plot of different priors  for 2 regressors  for
          $\deltav=(1,1)$ and $\deltav=(0,0)$: Dirac/i-slab (left),
          Dirac/g-slab (middle), Dirac/f-slab
          (right)\label{fig:pri}}
\end{figure}

$\Xv_\deltav$ is the design matrix consisting only of those columns of $\Xv$ corresponding to non-zero effects, i.e.~where $\delta_j = 1$. The g-slab is Zellner's g-prior \cite{zel:ass} for these effects and the f-slab is the corresponding fractional prior \cite{oha:fra}. The idea of the fractional prior is to use a fraction $b$ of the likelihood to determine a prior distribution for the parameters. In our specification the f-slab is not a fraction of the whole likelihood, but only of the part containing information on the regression coefficients $\alphav$. Note that in contrast to the i-slab, regression coefficients $\alpha_j$ are not independent conditional on $\deltav$ for g- and f-slab, where the joint distribution of all effects  with $\delta_j = 1$ is specified with a variance-covariance matrix equal to a scalar multiple of the Fisher information matrix. However, their mean is different: the g-slab is centered at the null vector, whereas the mean of f-slab is the LS estimate of the regression effects with $\delta_j=1$. Figure \ref{fig:pri} illustrates the differences between the three priors for two regressors showing the contours for the slab component for $\deltav = (1, 1)$ together with the position of the  spike for $\deltav = (0, 0)$.

\section{Inference}\label{sec:MCMC}

For both types of spike and slab priors posterior inference is feasible using MCMC methods, where the model parameters $(\mu, \deltav, \alphav, \omega, \verror)$ and additionally, under the NMIG prior, the scale parameters $\psiv = (\psi_1, \dots, \psi_d)$ are sampled from their conditional posteriors. Depending on the type of the spike component, different sampling schemes have to be used: Whereas for an absolutely continuous spike the indicators $\delta_j$ can be sampled conditionally on the effects $\alpha_j$, for a Dirac spike it is essential to draw $\deltav$ from the marginal posterior $p(\deltav|\mathbf{y})$ integrating over the parameters subject to selection, see \citeA{gew:var} and \citeA{smi-koh:non}. This requires evaluation of marginal likelihoods in each MCMC iteration. In  normal regression models with conjugate priors (which are used here) analytical integration over the regression effects is feasible and hence  marginal likelihoods can be computed  rather cheaply. Details of the MCMC sampling schemes are given in the following two subsections.

\subsection{MCMC for Absolutely Continuous Spikes}\label{sec:mcmc_con}

For priors with an absolutely continuous spike the full conditional distribution of $(\deltav, \psiv)$ is given as
$$
 p(\deltav, \psiv|\alphav, \omega, \mu, \verror, \yv)
 \propto
 \prod_{j=1}^d p(\alpha_j|\delta_j, \psi_j)
               p(\delta_j|\omega) p(\psi_j) p(\omega)
 \propto
 \prod_{j=1}^d p(\psi_j|\delta_j, \alpha_j)
               p(\delta_j|\alpha_j, \omega)\,.
$$
Therefore, $\deltav$ and $\psiv$ can be sampled together in one block and the sampling scheme involves the following steps:
\begin{enumerate}\vspace*{-2mm}
 \item[(1.)] Sample $\mu$ from its posterior $\mu|\verror, \yv \sim \Normal{\bar{y}, \verror/N}$.\vspace*{-2mm}
 \item[(2.)] Sample $\deltav$ and $\psiv$:\vspace*{-2mm}
      \begin{enumerate}
      \item [(2a.)]For $j = 1, \dots, d$ sample $\delta_j$ from
            $$
            p(\delta_{j} = 1|\alpha_j, \omega)
                         = \frac 1{1 + \displaystyle{\frac{1-\omega}{\omega}}L_j}\,,
            \qquad
            L_j = \frac{p_{\spike}(\alpha_j)}{p_{\slab}(\alpha_j)}\,.
            $$\vspace*{-2mm}
      \item[(2b.)] For normal spikes and slabs, set $\psi_j \equiv V$.
           For student spikes and slabs, where $\psi_j \sim
           \Gammainv{\nu, Q}$, sample $\psi_j$ from its conditional
           posterior
           $$
           \psi_j|\delta_j, \alpha_j \sim \Gammainv{\nu + \frac 12,
           Q + \frac{\alpha_j^2}{2 r(\delta_j)}}\,.
           $$
      \end{enumerate}\vspace*{-2mm}
 \item[(3.)] Sample $\omega$ from $\omega \sim \Betadis{a_{\omega} + d_1, b_{\omega} + d - d_1}$ where $d_1 = \sum\delta_j$. \vspace*{-2mm}
 \item[(4.)] Sample $\alphav$ from the normal posterior $\Normal{\av_N, \Av_N}$ where $\Av_N^{-1} = \frac 1{\verror}(\Xv'\Xv) + \Dv^{-1}$ and $\av_N = \Av_N\Xv'\yv_c/\verror$. $\Dv$ is a diagonal matrix with entries $r(\delta_j)\psi_j$, $j = 1, \dots, d$. \vspace*{-2mm}
 \item[(5.)] Sample the error variance $\verror$ from the posterior $\verror|\yv_c, \alphav \sim \Gammainv{s_N, S_N}$, where $s_N = (N-1)/2$ and $S_N = \tfrac 12(\yv_c-\Xv\alphav)'(\yv_c-\Xv\alphav)$.
\end{enumerate}

\subsection{Sampling Steps for a Dirac Spike} \label{sec:mcmc_dir}

For a Dirac spike, $\delta_j = 0$ implies $\alpha_j = 0$ and vice versa. To avoid reducibility of the Markov chain, it is essential to draw $\deltav$ from the marginal posterior
$$
 p(\deltav|\yv) \propto p(\yv|\deltav)p(\deltav)\,,
$$
where effects subject to selection are integrated out. Here $p(\yv|\deltav)$ denotes the marginal likelihood of the linear regression model (\ref{regmod}) with design matrix $\Xv_\deltav$. For Dirac spikes combined with i-, g- or f-slab on
$\alphav_\deltav$ the marginal likelihood can be derived analytically as
\begin{equation}\label{marlik1}
 p(\yv|\deltav)
 = \frac{1}{\sqrt{N}(2\pi)^{(N-1)/2}}
   \frac{|\Av_\deltav|^{1/2}}{|\Av_{0,\deltav}|^{1/2}}
   \frac{\Gamma(s_N)}{S_N^{s_N}}\,,
\end{equation}
where  $s_N = (N-1)/2$ and $S_N = \frac 12(\yv_c'\yv_c - \av_\deltav' \Av_\deltav^{-1} \av_\deltav)$. $\av_{\deltav}$ and $\Av_{\deltav}$ are parameters of the  posterior of $\alphav_\deltav$:  $\Av_{\deltav} = ((\Xv_{\deltav}'\Xv_\deltav) + \frac 1c\mbI)^{-1}$ for the i-slab, $\Av_{\deltav} = \frac{g}{g+1} (\Xv_{\deltav}' \Xv_\deltav)^{-1}$ for the g-slab and $\Av_{\deltav} = (\Xv_{\deltav}' \Xv_\deltav)^{-1}$ for the f-slab; the posterior mean is $\av_{\deltav} = \Av_{\deltav} \Xv_\deltav'\yv_c$ for any of the three slabs. Details are given in Appendix \ref{app:ml}.

With this marginalization it is possible to sample the parameters $\deltav$, $\verror$ and $\mu$ in one block. Hence, the MCMC scheme for Dirac spikes  involves the following steps:
\begin{enumerate}\vspace*{-2mm}
 \item[(1.)] Sample $(\deltav,\verror,\mu)$ from the posterior $p(\deltav|\yv) p(\verror|\yv,\deltav)p(\mu|\yv,\deltav,\verror)$. \vspace*{-2mm}
 \begin{enumerate}
      \item[(1a.)] Sample each element $\delta_{j}$ of the indicator vector $\deltav$ separately from $p(\delta_{j}=1|\deltav_{\backslash j}, \yv)$ given as
           $$
           p(\delta_{j}=1|\deltav_{\backslash j}, \yv)
           =
           \frac 1{1 + \displaystyle{\frac{1-\omega}{\omega}}R_j}\,,
           \qquad
           R_j = \frac{p(\yv|\delta_j=0,\deltav_{\backslash j})}
                      {p(\yv|\delta_j=1,\deltav_{\backslash j})}\,.
           $$
           Here $\deltav_{\backslash j}$ denotes the vector $\deltav$ consisting of all elements of $\deltav$ except $\delta_j$. Elements of $\deltav$ are updated in a random permutation order.
           \vspace*{-2mm}
      \item[(1b.)] Sample the error variance $\verror$ from the $\Gammainv{s_N, S_N}$-distribution.\vspace*{-2mm}
      \item [(1c.)] Sample the mean $\mu$ from the $\Normal{\bar{y}, \verror/N}$-distribution.\vspace*{-2mm}
      \end{enumerate}
 \item[(2.)] Sample $\omega$ from $\omega \sim \Betadis{a_{\omega}+d_1, b_{\omega}+d-d_1}$, where $d_1=\sum \delta_j$.\vspace*{-2mm}
 \item[(3.)] Set $\alpha_j=0$ if $\delta_j=0$. Sample the  non-zero elements $\alphav_{\deltav}$ from  the normal posterior $\Normal{\av_{\deltav}, \Av_{\deltav}\verror}$.
 \end{enumerate}
For both g- and f-slab, the posterior variance covariance matrix $\Av_{\deltav}$ is a scalar multiple of the prior variance covariance matrix $\Av_{0, \deltav}$. Thus  for  computing the marginal likelihood (\ref{marlik1}), the determinant of $\Av_{\deltav}$ is not required  which speeds up sampling  compared to i-slabs.

\section{Simulated Data} \label{sec:Simu}

We investigate performance of the different MCMC implementations for simulated data. Interest lies in correct selection of regressors as well as sampling efficiency of posterior inclusion probabilities. We expect draws of the posterior probabilities
$p^{(m)}(\delta_j=1)$, $m = 1, \dots, M$ to have higher autocorrelations for continuous than for Dirac spikes. It is however not obvious which implementation will have higher computational cost in CPU time: With a Dirac spike only coefficients with $\delta_j=1$ have to be sampled, as those with $\delta_j=0$ are restricted exactly to zero, whereas for a continuous spike the dimension of the model is not reduced during MCMC. On the other hand, specifying a continuous spike will save CPU time as no marginal likelihoods have to be computed.

To investigate correct model selection we simulate 100 data sets with $N = 40$ observations from a linear regression model with mean $\mu = 1$ and $\verror=1$ and nine covariates. We consider two setups for the covariate vectors $\xv_j$, which are drawn from a $\Normal{\zerov, \Cv}$-distribution: independent regressors, where $\Cv = \mbI$, and correlated regressors generated as in \citeA{tib:reg}, where $\Cv$ is a correlation matrix with $C_{jk} = \rho^{|j-k|}$ with $\rho = 0.8$. For both independent and correlated regressors we set three regression effects to each of the values ``2" (strong effects), ``0.2" (weak effects) and ``0" (zero effects).

In the simulation studies, we use an uninformative $\Betadis{1, 1}$-prior for $\omega$. To mimic Dirac spikes closely, a small variance ratio $r$ of continuous spikes and slabs would be preferred, however $r$ should not be too small to avoid MCMC getting stuck in the spike component. Following the recommendations in \citeA{geo-mcc:var} we set $r = 1/10000$.

It is well known that the choice of the slab distribution is critical for model selection. Our choice for the slab variance is motivated by the fact that model selection based on Bayes factors is consistent for the g-prior with $g = N$ (see \citeNP{fer-etal:ben}). Hence we choose $g = N$ and match the variances of the other slabs to equal the variance of the g-slab if  regressors are orthogonal, i.e.\ we
choose $b = 1/N$ and $V = 1$. For the NMIG-prior we choose $\nu = 5$, which corresponds to a t-distribution with 10 degrees of freedom, and $Q = 4$.

For each data set, MCMC was run for $M = 5000$ iterations after a burn-in of 1000 draws. The first 500 draws of the burn-in were drawn from the full model including all regressors.

\subsection{Model Selection Performance}\label{sec:selres}

\begin{figure}[h!]
 \includegraphics[width=15cm]{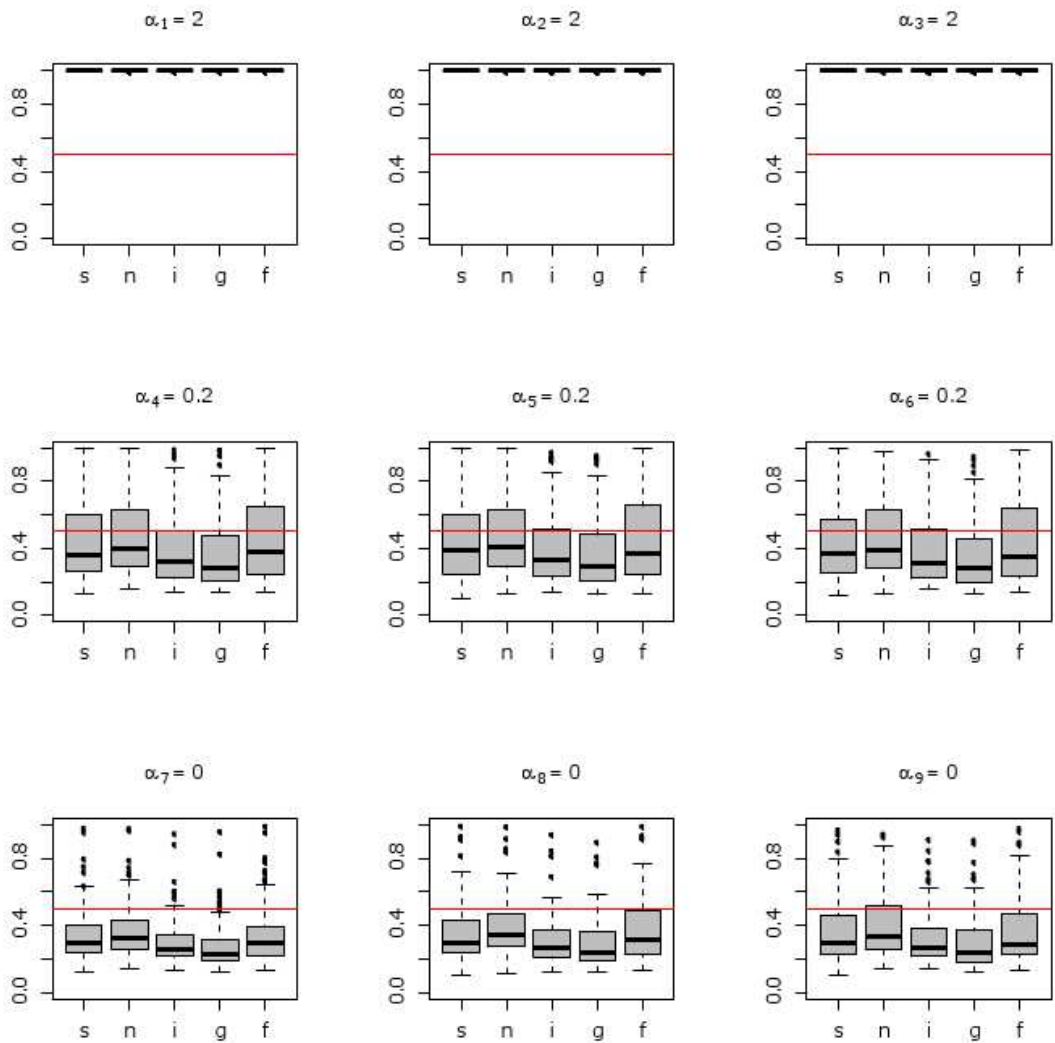}
 \caption{Independent regressors: Posterior inclusion probabilities
          of each regressor for 100 simulated data sets (s=SSVS prior,
          n=NMIG prior, i=Dirac/i-slab, g=Dirac/g-slab, f=Dirac/f-slab)
          \label{fig:ind_res}}
\end{figure}

Posterior inclusion probabilities are estimated by their posterior mean, i.e.\ the average of the inclusion probabilities $p^{(m)}(\delta_j=1)$ in the MCMC iterations. Figure \ref{fig:ind_res} shows box-plots of these estimates in the 100 simulated data sets with independent regressors. Regressors with strong effect are perfectly classified with estimated posterior inclusion probabilities being equal to 1 (rounded) in all 100 data sets. Variation of the estimated posterior inclusion probabilities is high for regressors with weak and zero effect which indicates that regression coefficients of smaller magnitude are hard to classify. Posterior inclusion probabilities tend to be slightly smaller for the Dirac/i-slab and the Dirac/g-slab priors than for the other priors.

For orthogonal regressors \citeA{bar-ber:opt} showed that the median probability model, i.e.\ the model including regressors with posterior inclusion probability larger than 0.5, is the best model with regard to predictive performance. Table
\ref{tab:ind_res} reports the number of data sets where each of the regressors with weak or zero effect is included in the median probability model. Results are not shown for regressors with strong effect as these are included in all 100 data sets under any prior. Whereas under the Dirac spike combined with i- or g-slab classification is better for zero effects, weak effects are detected less often than
under the other three priors. Overall performance is similar for all priors with mean misclassification rates (computed over weak and zeros effects) from 41.6 \% (Dirac/f-slab) to 43.3 \% (NMIG).
\begin{table}
 \begin{center}
 \begin{small}
 \caption{Independent regressors: Number of data sets where
          $\hat p(\delta_j=1) > 0.5$.\label{tab:ind_res}}

 \vspace*{2mm}
 \begin{tabular}{|cr|cc|ccc|}
 \hline
 &
 & \multicolumn{2}{c|}{Continuous spike}
 & \multicolumn{3}{c|}{Dirac spike} \\
 $j$ & $\alpha_j$ & SSVS & NMIG & i-slab & g-slab & f-slab\\
 \hline
 4 & 0.2 & 31 & 36 & 25 & 23 & 36 \\
 5 & 0.2 & 33 & 35 & 26 & 25 & 37 \\
 6 & 0.2 & 28 & 32 & 26 & 23 & 38 \\
 7 & 0   & 12 & 15 & 11 &  9 & 15 \\
 8 & 0   & 18 & 22 & 11 &  8 & 24 \\
 9 & 0   & 21 & 26 & 13 & 11 & 22 \\
 \hline
 \end{tabular}
 \end{small}
 \end{center}
\end{table}

\begin{figure}[t!]
 \includegraphics[width=15cm]{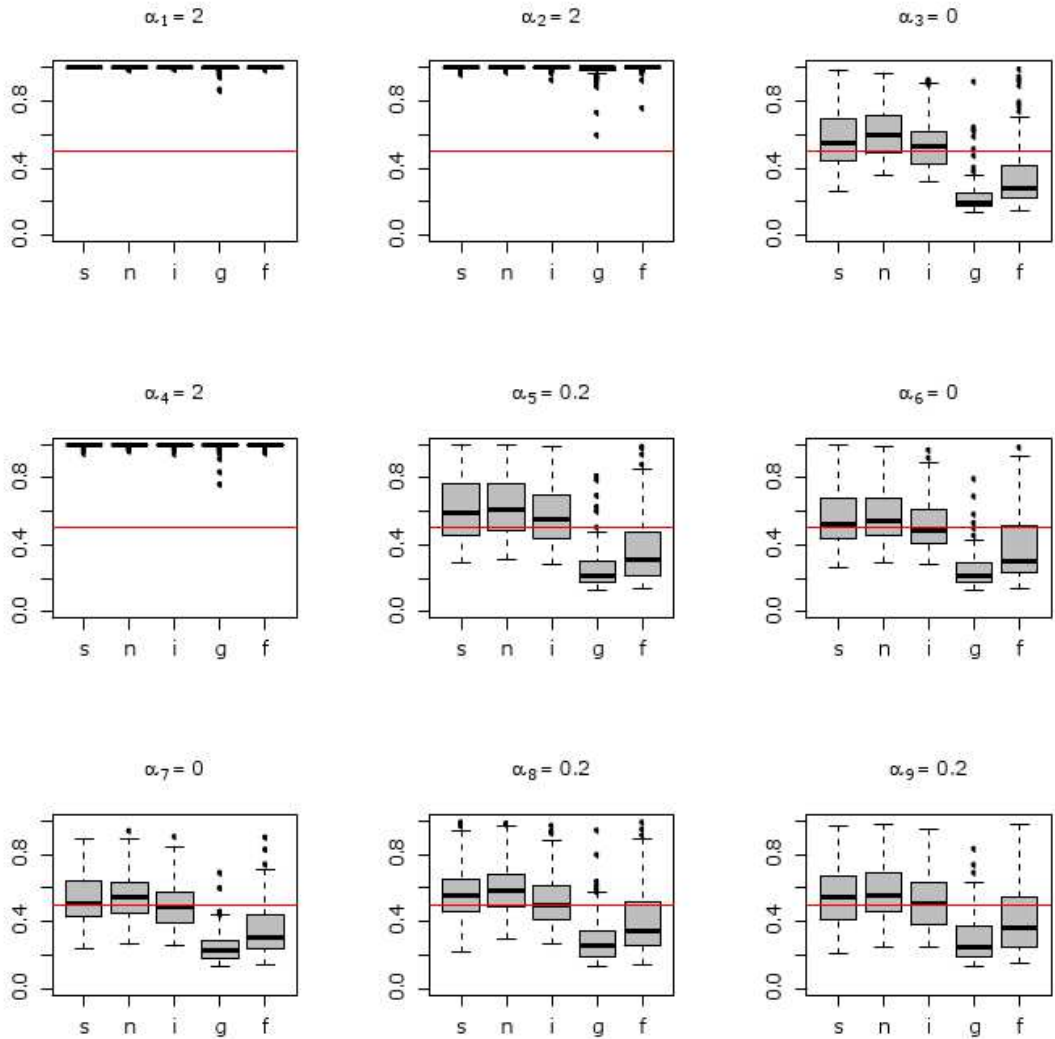}
 \caption{Correlated regressors: Posterior inclusion probabilities
          of each regressor for 100 simulated data sets (s=SSVS prior,
          n=NMIG prior, i=Dirac/i-slab, g=Dirac/g-slab, f=Dirac/f-slab).
          \label{fig:cor_res}}
\end{figure}

Figure \ref{fig:cor_res} shows the estimated posterior inclusion probabilities for simulated data with correlated regressors. The order of regressors with strong, weak and zero effects is different now, to get insight in the effects of correlations which are highest for neighboring regressors. Posterior inclusion probabilities of regressors with strong effects show  more variation than for independent regressors but are close to 1 in almost all cases. A pronounced difference however occurs  for regressors with weak and zero effects, which are slightly smaller for the Dirac/g-slab and Dirac/f-slab prior but considerably higher for priors with independent slabs (Dirac/i-slab, SSVS and NMIG) than in Figure \ref{fig:ind_res}. As a consequence, regressors with weak and zero effects are included in the median probability model less often under the Dirac/g-slab and Dirac/f-slab prior but more often under priors with independent slabs, when regressors are correlated. Table \ref{tab:cor_res} reports in how many data sets each regressor is included in the median probability model. As estimated posterior inclusion probabilities of regressors with strong effects are higher than 0.5 in all data sets (except in one data set for the g-slab), only results for weak and zero effects are given. Mean misclassification rates (computed over  weak and zeros effects) are higher than for independent regressors, but again very similar, ranging from 47.5~\% (Dirac/f-slab) to 48~\% (Dirac/i-slab). Obviously the correlation structure of the prior has an effect on the posterior inclusion probability when regressors are correlated. We will return to this issue in Section \ref{sec:theo_corr}, where we investigate this
effect analytically, though in a simpler setting.

\begin{table}[ht]
 \begin{center}
 \begin{small}
 \caption{Correlated regressors: Number of data sets where
          $\hat p(\delta_j=1) > 0.5$\label{tab:cor_res}}

 \vspace*{2mm}
 \begin{tabular}{|cr|cc|ccc|}
 \hline
 &
 & \multicolumn{2}{c|}{Continuous spike}
 & \multicolumn{3}{c|}{Dirac spike} \\
 $j$ & $\alpha_j$ & SSVS & NMIG & i-slab & g-slab & f-slab\\
 \hline
 3 & 0   & 62 & 66 & 58 &  6 & 19 \\
 5 & 0.2 & 66 & 73 & 60 &  6 & 22 \\
 6 & 0   & 60 & 66 & 44 &  5 & 26 \\
 7 & 0   & 55 & 63 & 48 &  2 & 18 \\
 8 & 0.2 & 67 & 73 & 50 & 10 & 26 \\
 9 & 0.2 & 57 & 63 & 52 & 10 & 30 \\
 \hline
 \end{tabular}
 \end{small}
 \end{center}
\end{table}

Further simulations carried out in \citeA{mal:bay} indicate that posterior inclusion probabilities depend on the variance of the slab component: Posterior inclusion probabilities decrease with increasing slab variance. This is another issue which we
investigate analytically in Section \ref{sec:theo} and illustrate in the application in Section \ref{sec:App}.

\subsection{Comparing Sampling Efficiencies}

As MCMC draws are correlated, it is of interest to compare MCMC implementations for the different priors with respect to their sampling efficiency. Table \ref{tab:ineff} reports mean inefficiency factors (also called integrated autocorrelation times) for regressors with weak and zero effects. Inefficiency factors, defined as $\tau = 1 + 2\sum_{l=1}^L \rho(l)$, where $\rho(l)$ is the empirical autocorrelation at lag $l$, were computed using the initial monotone sequence estimator \cite{gey:pra} for $L$. If inclusion probabilities $p^{(m)}(\delta_j=1)$ are numerically equal to 1 in all
iterations, inefficiency factors cannot be computed. This occurred for one effect in one data set under the SSVS prior and hence the average reported in Table \ref{tab:ineff} for the SSVS prior is based only on the remaining posterior inclusion probabilities.
\begin{table}[t]
 \begin{center}
 \begin{small}
 \caption{Averaged inefficiency factors\label{tab:ineff}}

 \vspace*{2mm}
 \begin{tabular}{|l|cc|ccc|}
 \hline
 & \multicolumn{2}{c|}{Continuous spike}
 & \multicolumn{3}{c|}{Dirac spike} \\
 Regressors  & SSVS & NMIG  & i-slab & g-slab & f-slab \\
 \hline
 Independent & 26.3 & 23.7  & 3.3    & 3.1    & 3.2 \\
 Correlated  & 30.1 & 27.2  & 3.7    & 2.5    & 2.9 \\
 \hline
 \end{tabular}
 \end{small}
 \end{center}
\end{table}
Interestingly for correlated regressors inefficiency factors are lower for the Dirac/g-slab and Dirac/f-slab prior and higher for priors with independent slab. This might result from the decrease/increase of posterior inclusion probabilities:
\citeA{wag-dul:bay} also observed smaller inefficiency factor for low inclusion probabilities, though in logit models.

As expected, inefficiency factors are considerably higher for priors with continuous spikes than for Dirac spikes. Further simulations in \citeA{mal:bay} showed that, for continuous spikes, the choice of the variance ratio $r$ as well as the actual implementation can have an impact on sampling efficiency: Autocorrelations and inefficiency are lower for higher values of $r$, e.g.\ $r = 1/1000$ yields similar estimates for posterior inclusion probabilities but with less autocorrelated draws. Under the NMIG prior, posterior inclusion probabilities could be computed alternatively conditional on the variance parameters $\psi_j$ as in \citeA{kon-etal:bay}, which however leads to considerably higher autocorrelations than using the marginal t-distribution.

To assess sampling efficiency with computing time taken into account, Table \ref{tab:ess} reports effective sample sizes per second averaged over weak and zero effects. The effective sample size  $ESS=M/\tau$ estimates the number of independent samples required to obtain  a parameter estimate with the same precision as the MCMC estimate. Results in Table \ref{tab:ess} are based on all MCMC chains, where inefficiency factors could be computed.
\begin{table}[t]
 \begin{center}
 \begin{small}
 \caption{Averaged effective sample size per sec.\label{tab:ess}}

 \vspace*{2mm}
 \begin{tabular}{|l|cc|ccc|}
 \hline
 & \multicolumn{2}{c|}{Continuous spike}
 & \multicolumn{3}{c|}{Dirac spike} \\
 Regressors  & SSVS & NMIG & i-slab & g-slab & f-slab \\
 \hline
 Independent & 33.3 & 27.6 & 16.6   & 34.3   & 23.2 \\
 Correlated  & 25.1 & 18.9 & 14.9   & 43.3   & 27.1 \\
 \hline
 \end{tabular}
 \end{small}
 \end{center}
\end{table}
Though sampling efficiency is much higher for priors with Dirac spikes differences in effective sample sizes are much less pronounced and priors with absolutely continuous spikes perform roughly similar to Dirac/g- and Dirac/f-slab. Even in this rather low-dimensional model with only nine regressors, computational cost for the Dirac/i-slab prior is too high to be outweighed by the smaller inefficiency factors. Due to lower inefficiency factors priors with g- and f-slabs have even higher $ESS/sec$ for correlated regressors.

\section{Posterior Inclusion Probabilities}\label{sec:theo}

Results of the simulation study indicate that posterior inclusion probabilities largely depend on the slab distribution. To get further insight into the effect of different slabs we investigate the inclusion probability of one regressor $\xv_j$ conditional on $\deltav_{\backslash j}$ for priors with Dirac spikes (i.e.\ the Dirac/i-slab, Dirac/g-slab and Dirac/f-slab prior) in two simple special cases: for orthogonal regressors and in a model with only two correlated regressors. For simplicity we assume that the error variance $\verror$ is known. Details on the computation of posterior inclusion probabilities are given in  Appendix \ref{app:post_incl}. We will denote by $s^2_y = \frac 1N \yv_c'\yv_c$ the sample variance of $\yv$, by $r_{yj}$ the sample correlation between $\yv$ and $\xv_j$ and by $s^2_j = \frac 1N \xv_j'\xv_j$ the sample variance of covariate $\xv_j$.

\subsection{Orthogonal Regressors}

\begin{figure}[t!]
 \includegraphics[width=7.5cm]{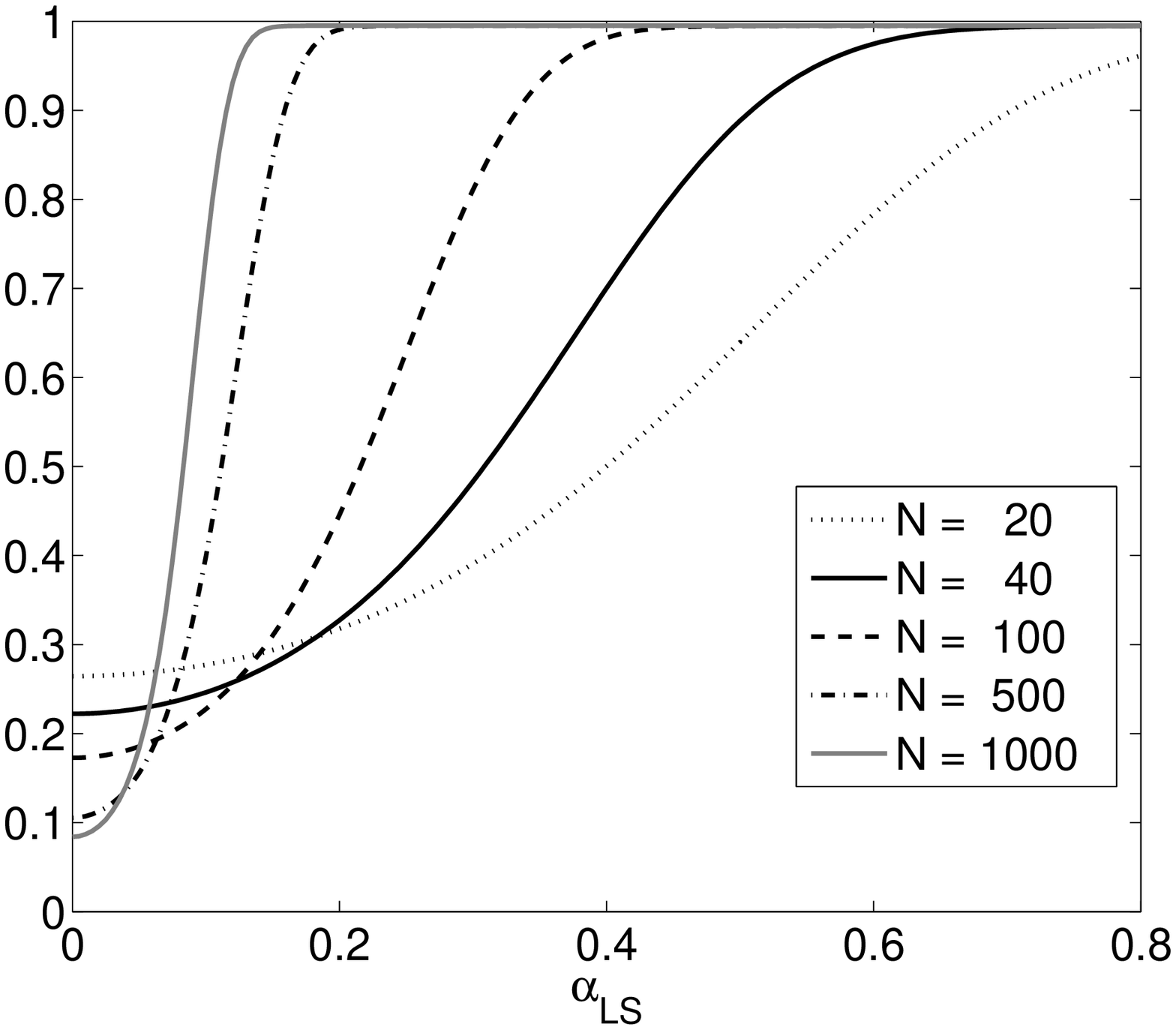}\hfill
 \includegraphics[width=7.5cm]{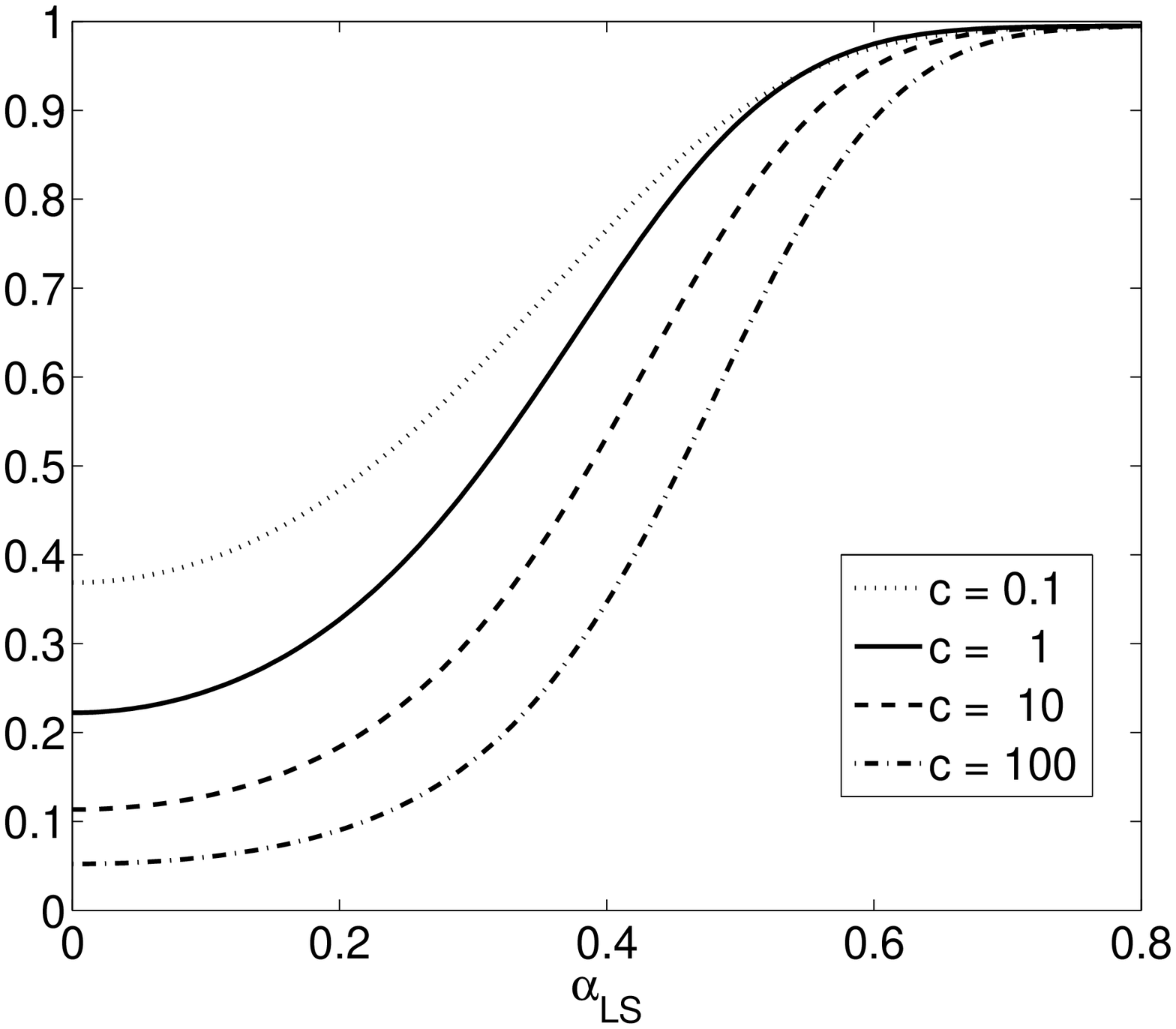}
 \caption{Independent regressors: Posterior inclusion probability
          under the Dirac/i-slab prior (for $\sigma^2=1$, integrated
          over $\omega$). Left: $c=1$, different values of $N$; right:
          $N=40$, different values of $c$.\label{fig:pinc}}
\end{figure}

For orthogonal regressors, i.e.\ $\Xv'\Xv = \diag(N s^2_j)$, $j = 1, \dots, d$, the posterior inclusion probability of $\xv_j$ can be written as a function of the LS-estimate $\hat\alpha_j=\frac {r_{yj} s_y}{s_j}$ as
\begin{equation}\label{postprob}
 p(\delta_j=1|\yv, \deltav_{\backslash j},\verror)
 = \frac 1{1+\exp(h(\hat\alpha_j, \theta)/2)\displaystyle{\frac{(1-\omega)}{\omega}}}\,,
\end{equation}
where $\theta$ is the variance parameter of the slab distribution, i.e.\ $c$ for the i-slab, $g$ for the g-slab and $b$ for the f-slab. Under any of the three slabs, the inclusion probability does not depend on $\deltav_{\backslash j}$. In particular we
obtain (see Appendix \ref{app:orth})
\begin{align}
 \text{i-slab: }\quad h(\hat\alpha_j,c)
 & = -N\frac{\hat\alpha^2_js_j^2}{\verror}\, \frac 1{1+1/(Ns^2_j c)} +
     \log(Ns^2_jc+1)\,,\label{islab}\\
 \text{g-slab: }\quad h(\hat\alpha_j,g)
 & = -\frac{N\hat\alpha^2_js_j^2}{\verror}\frac g{g+1}+\log(g+1)\,,\label{gslab}\\
 \text{f-slab: }\quad h(\hat\alpha_j,b)
 & = -\frac{N\hat\alpha^2_js_j^2}{\verror}(1-b) - \log(b)\,.\label{fslab}
\end{align}
In formulas (\ref{islab}) -- (\ref{fslab}) the first term is proportional to $N
\frac{\hat\alpha^2_j}{\verror}$ which, following \citeA{dey-etal:ind}, can be interpreted as the signal of the regression coefficient contained in the data. Hence, posterior inclusion probabilities increase with both, sample size $N$ and the size of the estimated effect $|\hat\alpha^2_j|$. The second term can be interpreted as a penalty term: It increases with the slab variance, and hence the posterior inclusion probability decreases as a function of the slab variance. Figure \ref{fig:pinc}
shows posterior inclusion probabilities under the i-slab as a function of the LS estimate $\hat\alpha$ for various samples sizes $N$ and variances $c$. In contrast to i-slabs the penalty term does not depend on the scale of the regressor under g- and
f-slabs. For standardized orthogonal regressors ($s^2_j = 1$) posterior inclusion probabilities are identical for g-slab and i-slab when $Nc = g$ and slightly higher under the f-slab when $b=1/g$. This corresponds to the simulation results, see Figure \ref{fig:ind_res}.

\begin{figure}[t!]
 \includegraphics[width=7.5cm]{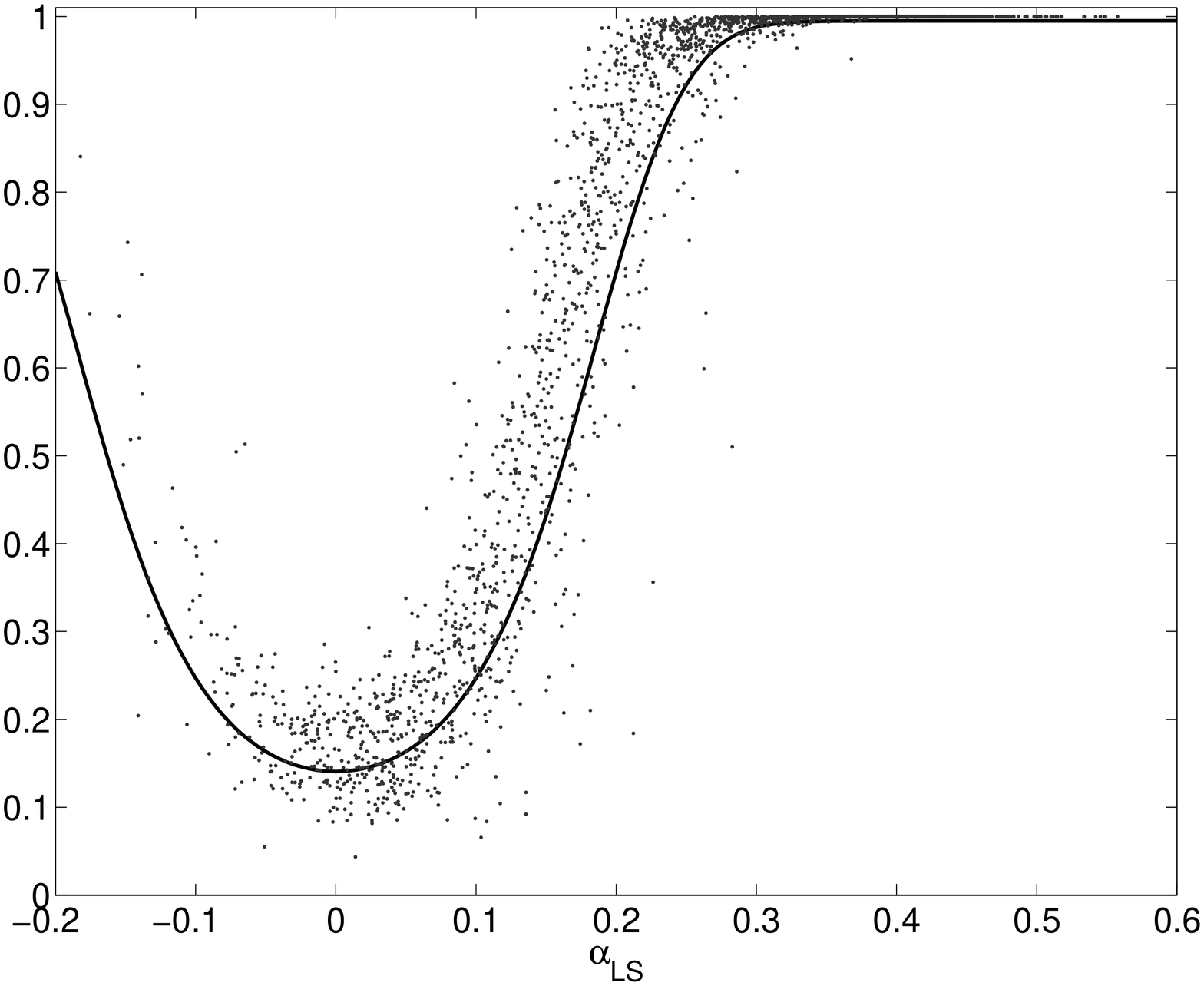}\hfill
 \includegraphics[width=7.5cm]{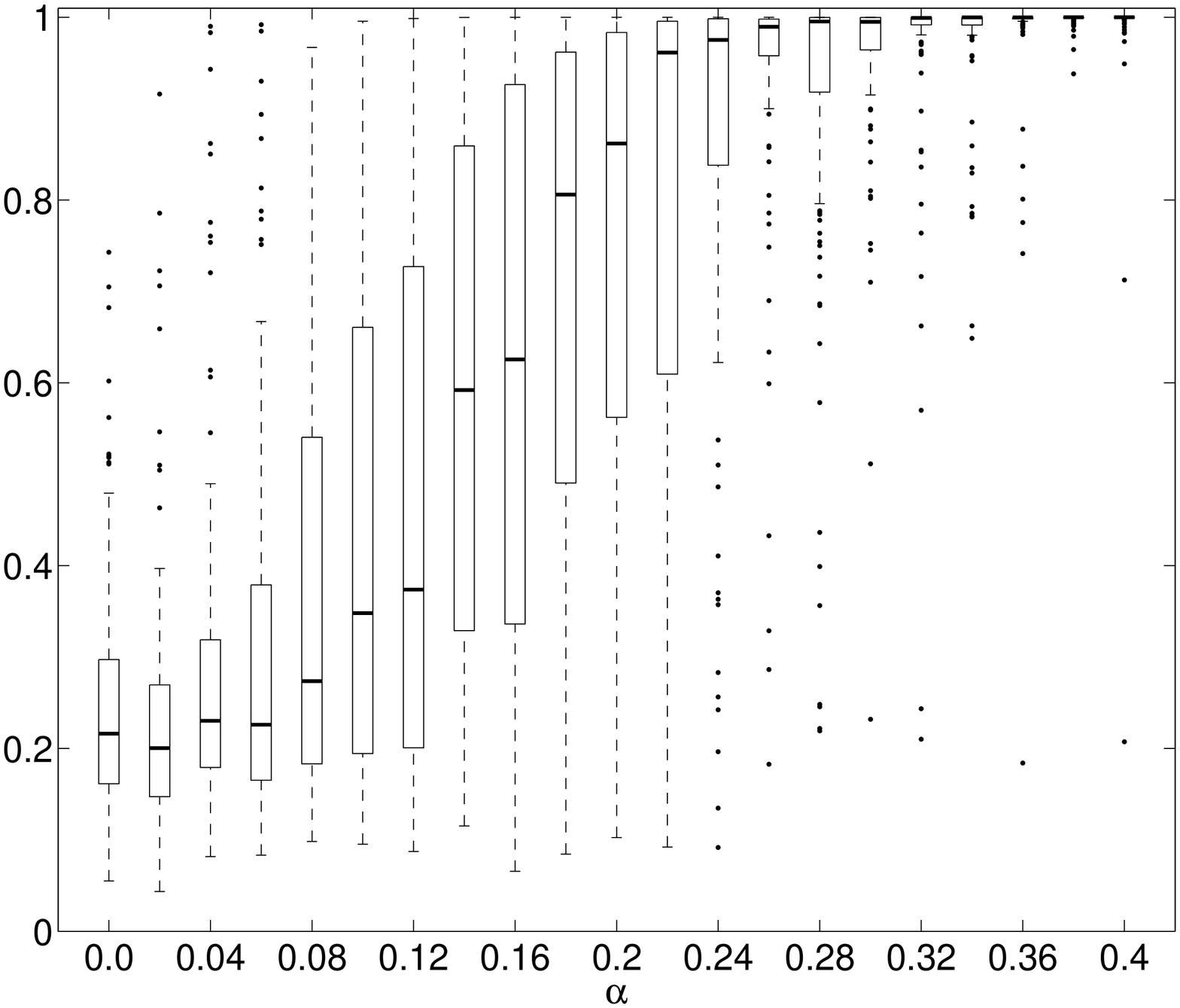}
 \caption{Simulated data: Posterior inclusion probabilities under
          Dirac/i-slab prior ($c=1$, $N=200$) as a function of the
          LS-estimate $\hat\alpha$ (left) and of the true effect
          $\alpha$ (right).\label{fig:simu2}}
\end{figure}

To illustrate the dependence of posterior inclusion probabilities on the effect signal $\hat{\alpha}$ we generated 100 data sets of size $N = 200$, with 21 regressors generated as independent standard normal random variables and effects from 0 to 0.4 in increments of 0.02. Posterior inclusion probabilities were estimated under the less restrictive assumption of unknown error variance using the MCMC scheme described in Section \ref{sec:mcmc_dir}. Figure \ref{fig:simu2} shows estimated posterior inclusion probabilities for the Dirac/i-slab plotted versus $\hat\alpha$ (left panel). Posterior inclusion probabilities do not exactly equal the theoretical values computed from formula (\ref{postprob}), which are shown as a line. This is not surprising as the assumptions  for the derivation of the formula are not met exactly: Firstly, due to stochastic variation regressors are not perfectly orthonormal and secondly, in the MCMC scheme the marginal likelihood  is computed using formula (\ref{marlik1}) with marginalization over the error variance $\verror$. In the right panel of Figure \ref{fig:simu2} estimated posterior inclusion probabilities are plotted against the ``true" effect sizes $\alpha$ used for data generation. Conditional on $\alpha$ variation of the posterior inclusion probabilities is much
higher as additionally the variation LS estimate $\hat{\alpha}$ is reflected.

\subsection{Two Correlated Regressors}\label{sec:theo_corr}

To investigate the effect of correlation between regressors we consider a model with only two standardized regressors $\xv_1$ and $\xv_2$ (i.e.\ $s_j^2=1$) and assume that $\xv_1$ is included in the model, i.e.\ $\delta_1=1$. We denote by $r_{12} = \frac 1N\xv_1'\xv_2$ the sample correlation between $\xv_1$ and $\xv_2$ and  by $\hat\alpha_2 = s_y(r_{y2} - r_{12}r_{y1})/(1 - r_{12}^2)$ the LS estimate of $\alpha_2$ in the model including both regressors.

\begin{figure}[t]
 \includegraphics[width=7.5cm]{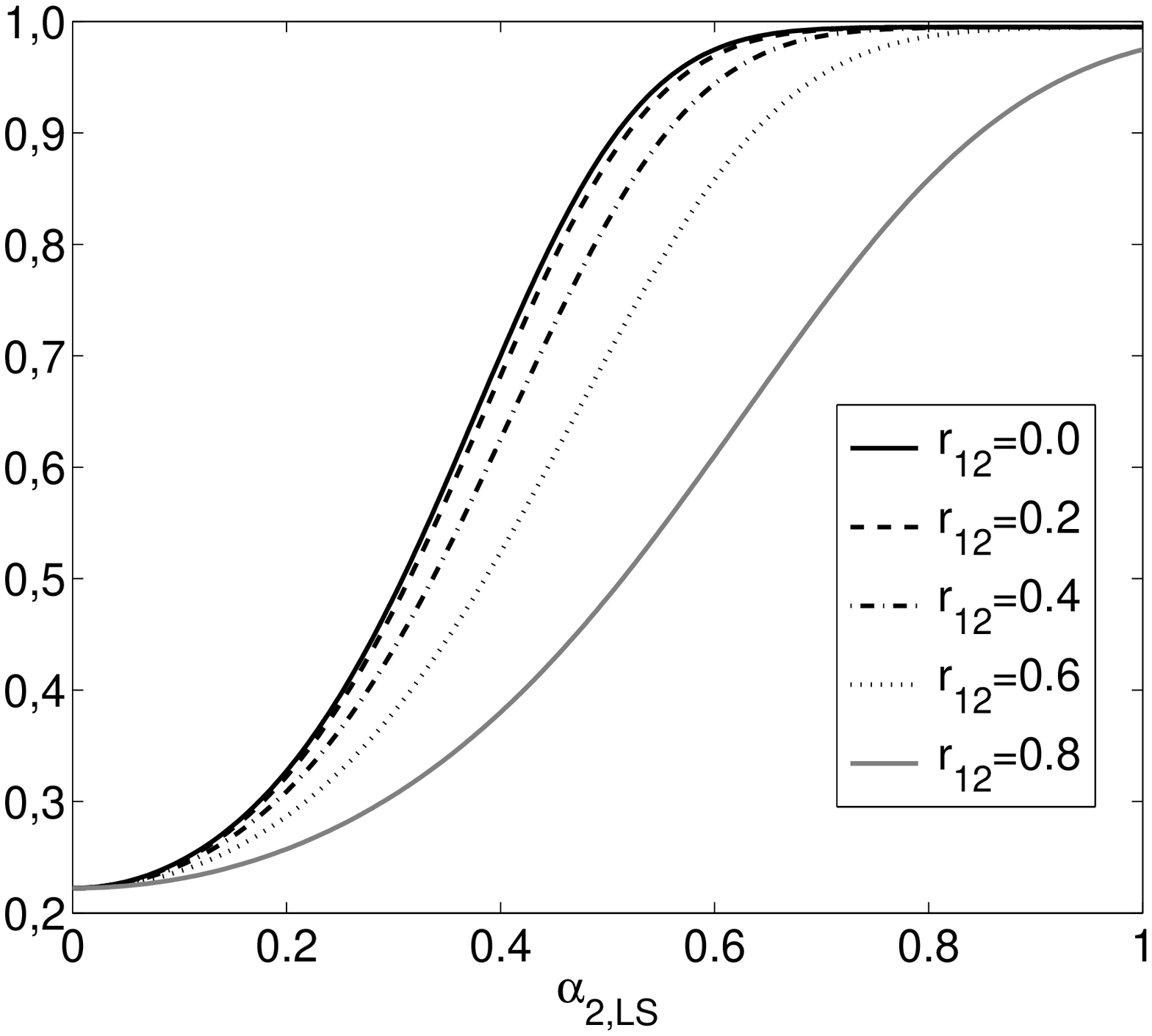}\hfill
 \includegraphics[width=7.5cm]{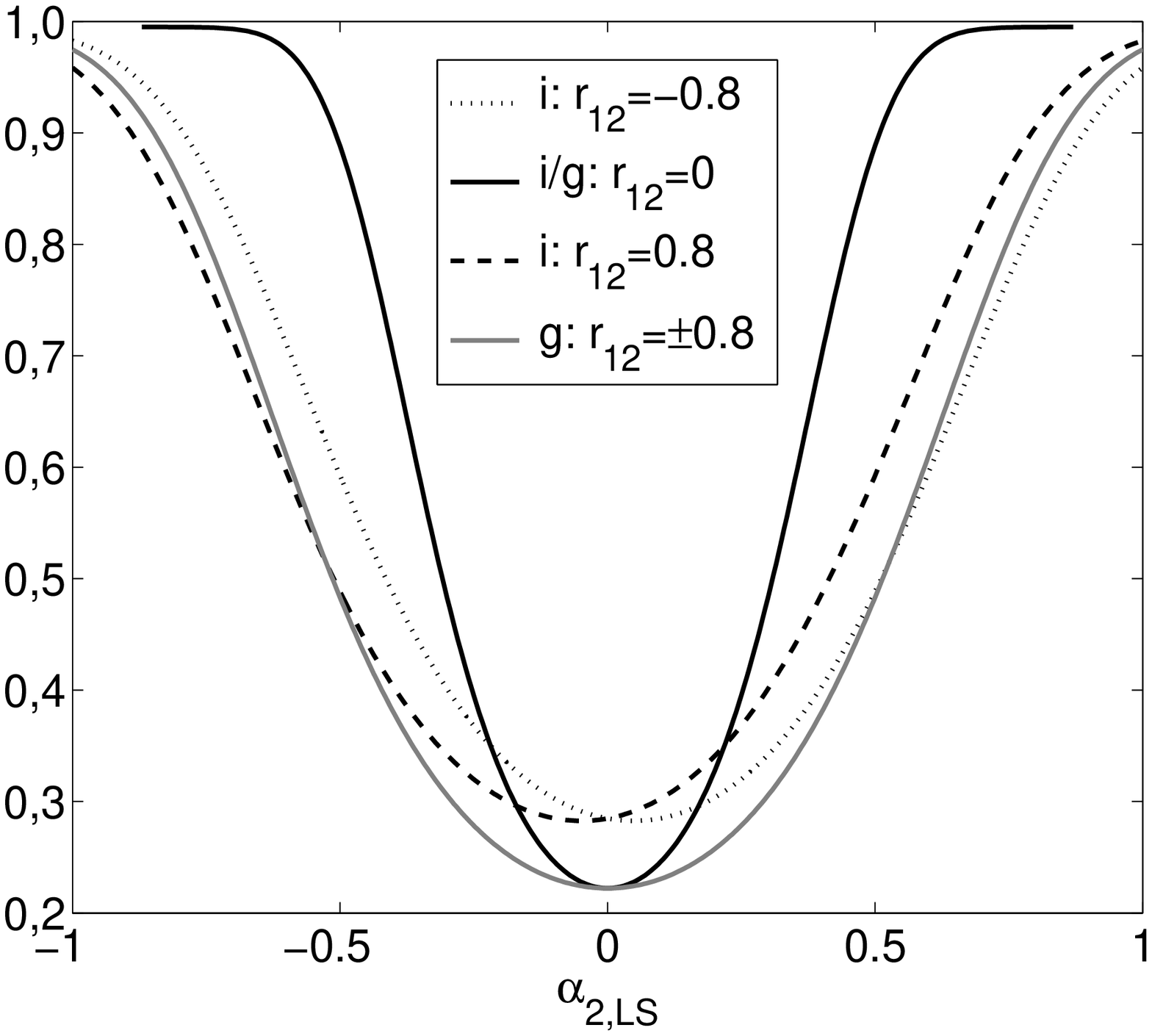}
 \caption{Correlated regressors: Posterior inclusion probability
          of regressor $\xv_2$, conditional on $\delta_1=1$ (integrated
          over $\omega$, $N=40$). Left: g-slab, different values of
          $r_{12}$, right: comparing g- and i-slab for different values
          of $r_{12}$ ($s_y=2$, $r_{1y}=0.9$).\label{fig:pinc_cor}}
\end{figure}

We are interested in the conditional posterior inclusion probability of $\xv_2$, which can be written as a function of
$$
 h(\hat\alpha_2, \sim)
 = 2\big(\log p(\yv|\delta_1=1, \delta_2=0, \verror)
       - \log p(\yv|\delta_1=1, \delta_2=1, \verror)\big)
$$
as in equation (\ref{postprob}). Under g- and f-slab it is straightforward to derive $h(\hat\alpha_2, \sim)$ as
\begin{align*}
 h(\hat\alpha_2,g) & = - \frac{N\hat\alpha_2^2}{\verror}(1-r_{12}^2)\,
 \frac g{g+1} + \log(g+1)\,,\\
 h(\hat\alpha_2,b) & = - \frac{N\hat\alpha_2^2}{\verror}(1-r_{12}^2)(1-b)
 - \log(b)\,,
\end{align*}
see Appendix \ref{app:corr} for details. For a given value of the LS estimate $\hat\alpha_2$, the probability of including $\xv_2$ additionally to $\xv_1$ in the model therefore decreases with the square of the correlation $r_{12}$ between the two regressors. Figure \ref{fig:pinc_cor} (left panel) shows the conditional posterior inclusion probabilities of $\xv_2$ under the Dirac/g-slab as a function of $\hat\alpha_2$ for different values of $r_{12}$. Obviously, for highly correlated regressors the inclusion probability of the second regressor can be reduced
dramatically.

For the Dirac/i-slab prior, simple but tedious algebra yields
$$
 h(\hat\alpha_2,c) = -\frac N{Q\verror}
 \Big(\hat\alpha_2(1-r_{12}^2) + \frac{r_{y2}s_y}{Nc}\Big)^2
 + \log\Big(Nc(1-r_{12}^2) + 1 + \frac{r_{12}^2}{1+1/(Nc)}\Big)\,,
$$
where
$$
 Q = (1-r_{12}^2) + \frac 1{Nc}(3-r_{12}^2) + \frac 3{(Nc)^2}
   + \frac 1{(Nc)^3}\,.
$$
The first summand in the function $h(\hat\alpha_2, c)$ is different from the corresponding term for g- and f-slab. However, as it is dominated by $-\frac{N\hat\alpha_2^2}{\verror}(1-r_{12}^2)$, this difference will vanish for increasing sample size $N$. Further, in contrast to g- and f-slab, the penalty term $\log(\sim)$ depends on the regressor correlation $r_{12}$ leading to less penalization of the additional regressor $\xv_2$ compared both to orthogonal
regressors and to g- and f-slabs. Therefore, posterior inclusion probabilities under i-slabs will be higher for correlated regressors. The conditional inclusion probabilities of $\xv_2$ under the i-slab depend not only on $\hat\alpha_2$, but also on $r_{2y}$ and are no longer symmetric in $\hat\alpha_2$, at least for small sample size $N$. This is shown in Figure \ref{fig:pinc_cor} (right panel), which compares the inclusion probability of $\xv_2$ for g- and i-slab for different correlations $r_{12}$. Posterior inclusion probabilities are considerably smaller under the g-slab for small absolute values of $\hat{\alpha}_2$. Results from our simulations, see Figure \ref{fig:cor_res}, suggest that differences in posterior inclusion probabilities under i- and g-slab can be even more pronounced in models with more regressors.

\section{Application} \label{sec:App}

We illustrate application of the different variable selection methods on a data set of psychiatric patients. Metabolic disorders and weight gain are common problems and side effects of psychiatric medication. To investigate how bodyweight and parameters of lipid and glucose metabolism are influenced by psychiatric inpatient treatment, a prospective study was performed at a department of the Wagner-Jauregg hospital in Upper Austria from October 2003 to March 2004. Several lipid and glucose parameters, namely total cholesterol ({\tt chol}), high density lipoprotein cholesterol ({\tt hdl}), low density lipoprotein cholesterol ({\tt ldl}), tri\-glycerides ({\tt nf}) and fasting glucose ({\tt nbz}) were measured at admission and at discharge of the department. Medication, if any, was assessed as prescribed at discharge and assigned to 16 drugs or types of drugs. Additionally, several patient-related variables were collected: age, sex, height, smoking, body mass index at admission and duration of the stay.

The focus of our analysis is to identify covariates influencing the {\tt change in HDL}, and we used the lipid and glucose values at admission, the 16 different drug types and all patient variables as potential regressors. Excluding  observations with missing values, leaves data on 231 patients with 27 regressors for the analysis. Pairwise correlations between covariates are smaller than 0.1 in most cases, only three correlations are higher than 0.4 ({\tt sex} and {\tt height}: $r=0.67$; {\tt chol\_admiss} and {\tt ldl\_admiss}: $r=0.86$ and {\tt drug A} and {\tt drug B}: $r = 0.89$).

\begin{figure}[t]
 \includegraphics[width=15cm]{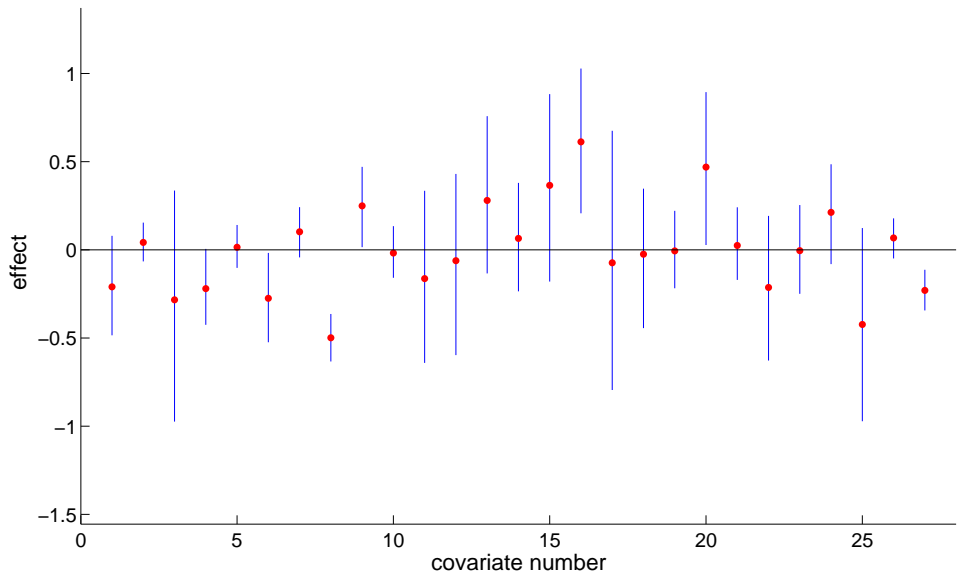}
 \caption{HDL data: Posterior means and 95\%- credible intervals
          for regression effects in the unrestricted model\label{fig:resun}}
\end{figure}

Following \citeA{gel-etal:wea}, metric covariates were standardized, and dummy covariates were centered. As a first step an exploratory Bayesian analysis of the unrestricted model under the prior $\Normal{\zerov, c\mbI}$ with $c = 5$  was carried out. Figure \ref{fig:resun} shows the posterior estimates and 95~\%-credible intervals of the regression effects. Only for 6 covariates (covariates number 6: {\tt chol\_admiss}, 8: {\tt hdl\_admiss}, 9: {\tt ldl\_admiss}, 16: {\tt drug F}, 20: {\tt drug J} and 27: {\tt bmi\_admiss}) these intervals do not contain zero, indicating that the corresponding effects ``significantly" differ from zero.

As a next step, MCMC for variable selection was run for $M = 5000$ iterations (after a burn-in of 1000, with the first 500 draws of the burn-in drawn from the unrestricted model) for Dirac spike priors and $M = 50000$ iterations (after 10000 burn-in with the first 5000 draws from the unrestricted model) for priors with absolutely continuous spikes. To match the slab variances the response was
standardized with the estimated residual standard deviation ($s = 15.4$) of the full regression model. Hyper-parameters were chosen as in the simulation studies: we used a variance ratio of $r = 1/10000$, $\nu = 5$ and $c = 1$ and the other parameters were set to $g = Nc$, $b = 1/g$, $V = c$ and $Q = 4$.

Posterior inclusion probabilities were roughly equal for all covariates under the Dirac/i-slab, the SSVS and  the NMIG prior, however, considerably smaller for Dirac/g- and Dirac/f-slab priors. Table \ref{tab:hdl_sel} reports estimated posterior inclusion probabilities for the covariates selected in the median probability model under the  Dirac/i-slab prior. Results correspond well with the exploratory analysis of  the unrestricted model: the selected covariates build a subset of those identified as having a ``significant'' effect, and in contrast to the exploratory
analysis, Bayesian variable selection automatically controls for multiple-testing.

From the medical point of view the goal of the analysis was to obtain a classification of covariates into those which have nearly zero effect and can be excluded from the model and others which potentially affect the response variable. Therefore, variable selection was not based on the Dirac/g- and f-slab-priors which more heavily penalize dependent regressors than independent slabs.
\begin{table}
 \begin{center}
 \begin{small}
 \caption{HDL data: Posterior inclusion probabilities (for $c=1$)
          \label{tab:hdl_sel}}

 \vspace*{2mm}
 \begin{tabular}{|l|cc|ccc|}
 \hline
 & \multicolumn{2}{c|}{Continuous spike}
 & \multicolumn{3}{c|}{Dirac spike} \\
 Covariate number & SSVS & NMIG & i-slab & g-slab & f-slab \\
 \hline
  8 (\tt hdl\_admiss) & 1.00 & 1.00 & 1.00 & 1.00 & 1.00 \\
 16 (\tt drug F)      & 0.78 & 0.82 & 0.81 & 0.49 & 0.49 \\
 20 (\tt drug J)      & 0.63 & 0.61 & 0.68 & 0.29 & 0.29 \\
 27 (\tt bmi\_admiss) & 0.56 & 0.53 & 0.62 & 0.34 & 0.32 \\
 \hline
 \end{tabular}
 \end{small}
 \end{center}
\end{table}

Table \ref{tab:hdl_ineff} shows inefficiency factors and effective sample size per sec.~ averaged over all covariates (except covariate 8). Again inefficiency factors of the posterior inclusion probabilities are considerably higher under priors with continuous spikes. However, when computational effort is taken into account again all priors except Dirac/i-slab prior perform similar.
\begin{table}[t]
 \begin{center}
 \begin{small}
 \caption{HDL data: Sampling efficiency of posterior inclusion
          probabilities\label{tab:hdl_ineff}}

 \vspace*{2mm}
 \begin{tabular}{|l|cc|crr|}
 \hline
 & \multicolumn{2}{c|}{Continuous spike}
 & \multicolumn{3}{c|}{Dirac spike} \\
 &  SSVS & NMIG & i-slab & g-slab & f-slab \\
 \hline
 Averaged inefficiency factor        & 57.7 & 43.2 & 3.1 &  2.1 &  2.3\\
 Averaged effective sample size/sec. & 15.6 & 12.8 & 5.9 & 16.9 & 10.9\\
  \hline
 \end{tabular}
 \end{small}
 \end{center}
\end{table}

\begin{figure}
 \includegraphics[width=7.5cm]{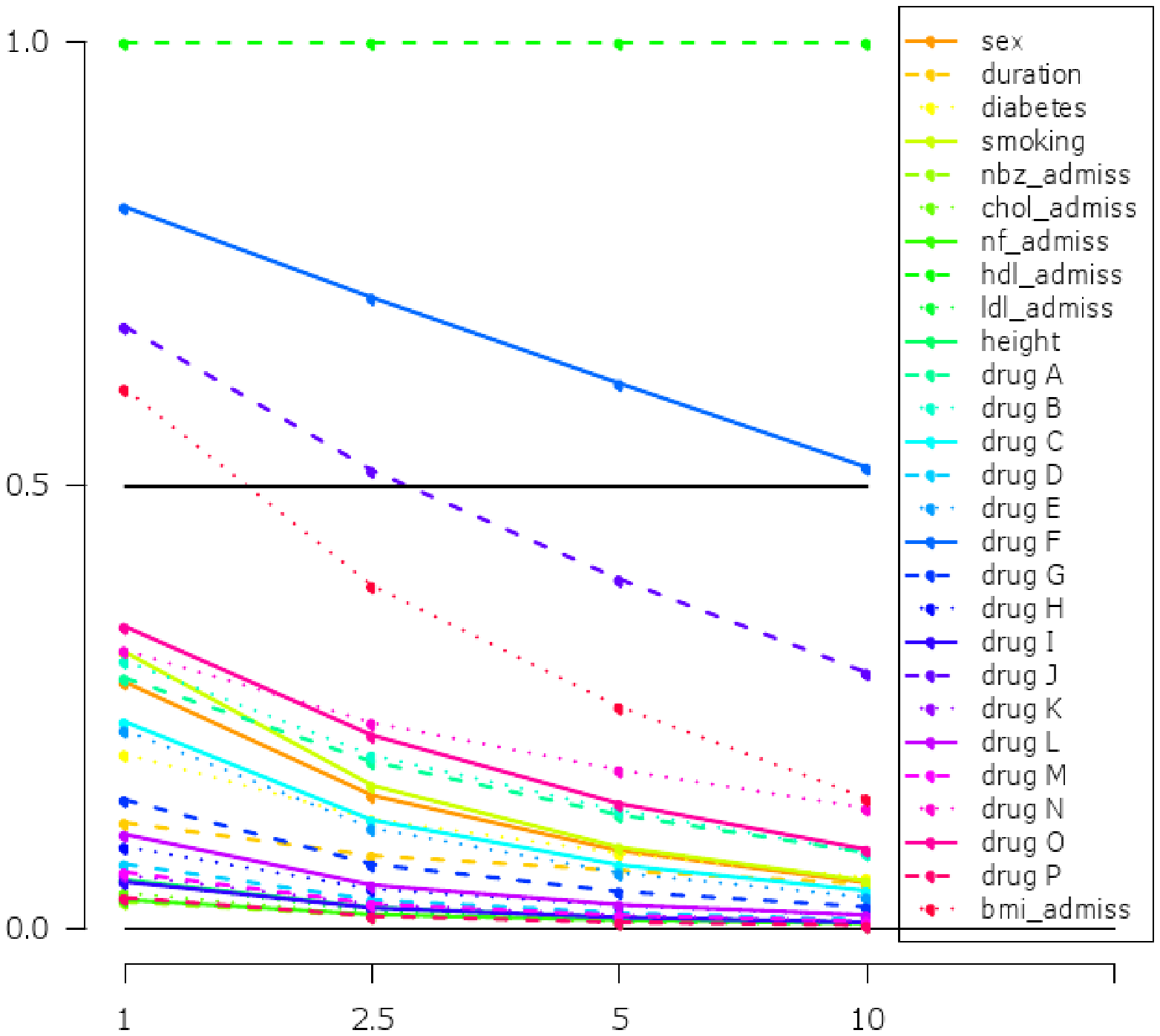}\hfill
 \includegraphics[width=7.5cm]{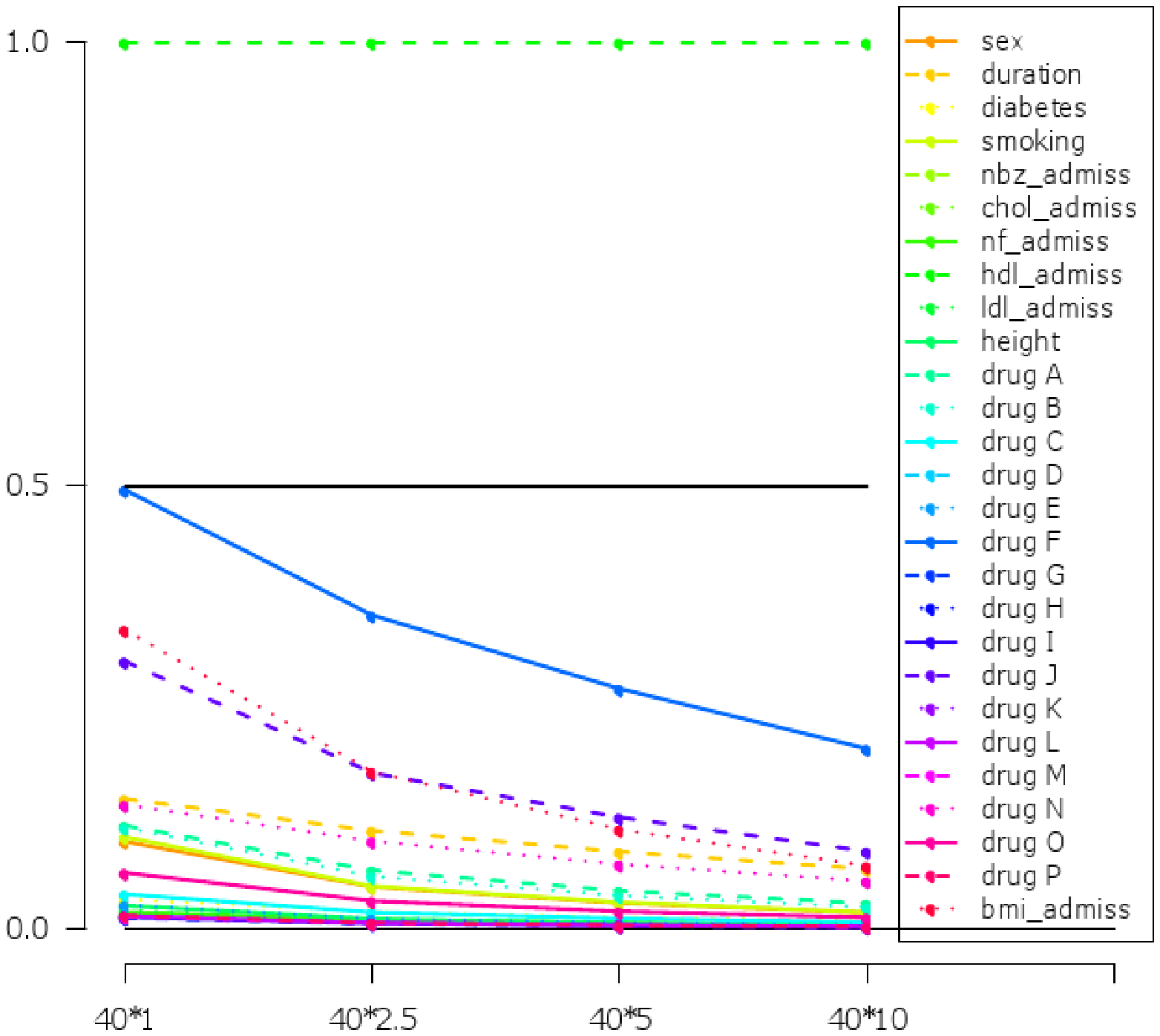}
 \caption{HDL data: Posterior inclusion probability paths for
          different slab variances $c$ for the Dirac/i-slab prior (left)
          and different values of $g$ for the Dirac/r-slab prior (right)
          \label{fig:path}}
\end{figure}

Finally, to study the effect of the slab variance, we ran MCMC for different values $c = 1, 2.5, 5, 10$ for the i-slab and corresponding parameters of the other priors. The resulting posterior inclusion probability paths shown in Figure \ref{fig:path} for the Dirac/i-slab and Dirac/g-slab priors, demonstrate the effect of increasing penalization of regressors for larger slab variances.

\section{Summary and Discussion}\label{sec:Sum}

We compared different spike and slab priors which are widely used for Bayesian variable selection. Simulation studies suggest that for orthogonal regressors different priors act rather similar when the slab variances are matched, which is confirmed by theoretical results for Dirac spike priors (and known  error variance). The posterior inclusion probability of a specific regressor increases with the signal of the effect in the data and decreases with the variance of the slab component. Compared to orthogonal regressors, both simulations as well as theoretical results, indicate that for a given effect signal in the data, posterior inclusion probabilities are smaller under g-and f-slabs and higher for priors with independent slabs if regressors are correlated. This result suggests to use g- or f-slabs in practical applications where interest is in  avoiding ``false positives" and independent slabs either with Dirac or continuous spikes if the goal is not to miss potentially important predictors.

From a computational point of view, priors with continuous spikes are a fast alternative to the Dirac/i-slab prior as higher autocorrelations are outweighed by less computation time. Mixing of the sampler is better for the NMIG than the SSVS prior at the cost of a small additional computational effort. MCMC getting stuck at $p(\delta_j=1) = 1$ is more severe for SSVS than NMIG priors, where it occurred only for regressors with strong effects.

A drawback of all priors considered here is that they do not well discriminate between regressors with zero and weak effects. Choosing a smaller variance for the slab component does not solve this problem as inclusion probabilities of all effects, even of zero effects, will increase. For Bayesian testing, \cite{joh-ros:use} recently proposed so called non-local prior densities, which are zero in the parameter space of the null hypothesis to facilitate separation between null and the alternative. Spike and slab priors compared in this paper could be modified in this direction with slab components having a mode different from zero. Prior information on the size of ``relevant" effects could be incorporated by specifying either one slab or, if no information on the effect sign is available, two slabs with a
positive and a  negative mode, respectively. For slabs which are normal or NMIG, MCMC schemes presented in this work could be used with slight modifications.

\subsection*{Acknowledgements}

The authors thank Univ.~Doz.~Prim.~Dr.~Hans Rittmannsberger (Wagner-Jauregg-Kranken\-haus Linz) for providing the data and many helpful comments. We would also like to thank the anonymous referee for his suggestions to improve the paper and Christoph Pamminger for careful reading of the manuscript.

\section*{Appendix}

\begin{appendix}
\section{Marginal Likelihoods}\label{app:ml}

We consider the normal regression model (\ref{regmod}) with $N \times d$ regressor matrix $\Xv$ with centered columns, i.e.~$\Xv' \mathbf{1} = \mathbf{0}$ with a prior of the structure
\begin{equation}\label{regprior}
 p(\mu, \verror, \alphav)
 \propto \frac 1{\verror}p(\alphav|\verror)\,.
\end{equation}
Integrating over $\mu$ we obtain
\begin{eqnarray*}
 p(\yv|\verror, \alphav, \Xv)
 &=& \int p(\yv|\verror, \mu, \alphav, \Xv)d\mu =\\
 &=& \frac 1{\sqrt{N}(2\pi\verror)^{(N-1)/2}}
   \exp\Big(-\frac 1{2 \verror} (\yv_c-\Xv\alphav)'(\yv_c-\Xv\alphav)\Big)\,,
\end{eqnarray*}
where $\yv_c = \yv - \mathbf{1} \bar y$. Further integration over $\alphav$ and $\sigma^2$ yields the conditional marginal likelihood $p(\yv|\verror, \Xv) = \int
p(\yv|\verror, \alphav, \Xv) p(\alphav|\verror)d\alphav$ and the marginal likelihood
$$
 p(\yv|\Xv) = \int p(\yv|\verror, \Xv)\frac 1{\verror} d\verror\,.
$$

\subsection{Conjugate Prior}

Under the conjugate prior $\alphav \sim \Normal{\av_0, \Av_0\verror}$ analytical integration is feasible, and the conditional marginal likelihood and marginal
likelihood are given as
\begin{eqnarray}
 p(\yv|\verror, \Xv)
 &=&
 \frac 1{\sqrt{N}(2\pi\verror)^{(N-1)/2}}
 \frac{|\Av_N|^{1/2}}{|\Av_0|^{1/2}}\exp\left(-\frac{S_N}{\verror}\right)
 \label{marlik_error}\\
 p(\yv|\Xv)
 &=&
 \frac 1{\sqrt{N}(2\pi)^{(N-1)/2}}
 \frac{|\Av_N|^{1/2}}{|\Av_0|^{1/2}}\frac{\Gamma(s_N)}{S_N^{s_N}}\label{marlik}\,.
\end{eqnarray}
Here $\av_N, \Av_N$ are the moments of the posterior distribution $p(\alphav|\verror, \yv)$:
$$
 \Av_N = \left(\Xv'\Xv + \Av_0^{-1}\right)^{-1}\,,
 \qquad
 \av_N = \Av_N\left(\Xv'\yv_c + \Av_0^{-1}\av_0\right)\,,
$$
and
$$
 S_N = \frac 12\left(\yv_c'\yv_c + \av_0'\Av_0^{-1}\av_0 - \av_N'\Av_N^{-1}\av_N\right)\,,
 \qquad
 s_N = \frac{N-1}2\,.
$$

Special cases are the independence prior $\alphav \sim \Normal{\zerov, c \mbI \verror}$ and the g-prior $\alphav \sim \mathcal{N}(\zerov$, $g(\Xv'\Xv)^{-1} \verror)$. In both cases $\av_N = \Av_N \Xv'\yv_c$ and hence $S_N$ simplifies to
$$
 S_N
 = \frac 12\left(\yv_c'\yv_c - \av_N'\Av_N^{-1}\av_N\right)
 = \frac 12\left(\yv_c'\yv_c - \yv_c'\Xv\Av_N\Xv'\yv_c\right)\,.
$$
For the independence prior, $|\Av_0| = c^d$ and $\Av_N = (\Xv'\Xv + \frac 1c\mbI)^{-1}$; for the g-prior $\Av_N = \frac g{g+1}(\Xv'\Xv)^{-1}$ and hence $|\Av_N|^{1/2}/|\Av_0|^{1/2} = (1+g)^{-d/2}$.

\subsection{Fractional Prior}

The fractional prior is obtained as a fraction of the likelihood, more specific we define the fractional prior as
$$
 p(\alphav|\verror)
 \propto p\left(\yv_c|\alphav, \verror\right)^b
 \propto \exp\Big(-\frac b{2\verror}(\yv_c-\Xv\alphav)'(\yv_c-\Xv\alphav)\Big)\,.
$$
The posterior, obtained by combining the prior with the remaining fraction of the likelihood, is the normal distribution with moments
$$
 \Av_N = (\Xv'\Xv)^{-1}\,,
 \qquad
 \av_N = (\Xv'\Xv)^{-1} \Xv'\yv_c\,.
$$
Conditional marginal likelihood and marginal likelihood can be computed from formulas (\ref{marlik_error}) and (\ref{marlik}) with $S_N =\frac 1{2} (1-b) \yv_c' (\mbI- \Xv (\Xv'\Xv)^{-1}\Xv')\yv_c$ and $|\Av_N|^{1/2}/|\Av_0|^{1/2} = b^{d/2}$.

\section{Posterior Inclusion Probabilities}\label{app:post_incl}

We compute posterior inclusion probabilities for a Dirac spike combined with i-, g- and f-slab. Without loss of generality, we consider posterior inclusion of last regressor $\xv_d$ conditional on $\deltav_{\backslash d}$. Further, we condition on $\verror$ and compute the posterior inclusion probability as
$$
 p(\delta_d=1|\yv, \deltav_{\backslash d}, \verror)
 =
 \frac 1{1 + \displaystyle{\frac{p(\yv|\deltav_{\backslash d}, \delta_d=0, \verror)}
                  {p(\yv|\delta_{\backslash d}, \delta_d=1, \verror)}
 \frac{(1-\omega)}{\omega}}}\,.
$$

We use the notation $\xv_j'\xv_j = N s^2_j$, $\yv_c'\xv_j = N s_j s_y r_{yj}$, $j = 1, \dots, d$ and $\yv_c'\yv_c = Ns^2_y$ and denote by $\hat\alpha_j = s_{yj}/s_j^2 = r_{yj}s_y/s_j$ the LS-estimator of $\alpha_j$. It will turn out that the conditional
posterior inclusion probability of regressor $\xv_d$ can be written as a function of $\hat\alpha_d$ and additional parameters $\theta$, depending on the slab, as
$$
 p(\delta_d=1|\yv, \deltav_{\backslash d}, \verror)
 =
 \frac 1{1 + \exp(h(\hat\alpha_d, \theta)/2)\displaystyle{\frac{(1-\omega)}{\omega}}}\,.
$$

\subsection{Orthogonal Regressors}\label{app:orth}

Let $\deltav^* = (\deltav_{\backslash d}, 1)$. For orthogonal regressors, both prior and posterior covariance matrix $\Av_{0, \deltav^*}$ and $\Av_{\deltav^*}$ are diagonal matrices for any of the priors on $\alphav_{\deltav^*}$ considered here. Denoting by $\Av_{\deltav^*, 0}(d)$, $\Av_{\deltav^*}(d)$, $\av_{0, \deltav^*}(d)$ and $\av_{\deltav^*}(d)$ the $d$-th element of $\Av_{0, \deltav^*}$, $\Av_{\deltav^*}$, $\av_{0, \deltav^*}$ and $\av_{\deltav^*}$, respectively, we obtain
\begin{align}\label{hfun1}
 h(\hat\alpha_d, \theta)
 & = 2\log\frac{p(\yv|\deltav_{\backslash d}, \delta_d=0, \verror)}
               {p(\yv|\deltav_{\backslash d}, \delta_d=1, \verror)}\\
             \label{hfun2}
 &= -\frac 1{\verror}
    \left(\frac{\big(\av_{\deltav^*}(d)\big)^2}{\Av_{\deltav^*}(d)}
        - \frac{\big(\av_{0, \deltav^*}(d)\big)^2}{\Av_{0, \deltav^*}(d)}\right)
  + \log\frac{\Av_{0, \deltav^*}(d)}{\Av_{\deltav^*}(d)}\,.
\end{align}
Further, under any of the three slabs,
$$
 \frac{\big(\av_{\deltav^*}(d)\big)^2}{\Av_{\deltav^*}(d)}
 =
 \frac{(\yv_c'\xv_d)^2}{1/\Av_{\deltav^*}(d)}
 =
 \frac{(N s_d s_y r_{yd})^2}{1/\Av_{\deltav^*}(d)}
 =
 \frac{(N s_d)^2 s_d^2\hat\alpha^2_d}{1/\Av_{\deltav^*}(d)}\,.
$$

For the i-slab with $\av_{0, \deltav^*}(d) = 0$, $\Av_{0, \deltav^*}(d) = c$ and $1/\Av_{\deltav^*}(d) = \xv_d'\xv_d+1/c = Ns_d^2 + 1/c$ we get
$$
 (\av_{\deltav^*}(d))^2/\Av_{\deltav^*}
 =
 N\hat\alpha^2_d s_d^2\,\frac 1{1 + 1/(N s_d^2 c)}\,.
$$
Thus, $h$ is a function of $\hat\alpha_d$ and $c$, given as
$$
 h(\hat\alpha_d, c) =
 -N\frac{\hat\alpha^2_ds_d^2}{\verror}\,\frac 1{1+1/(Ns^2_d c)}
 + \log(Ns^2_dc + 1)\,.
$$

For the g-slab, inserting $\av_{0, \deltav^*}(d) = 0$, $\Av_{0, \deltav^*}(d) = g/(N s^2_d)$ and $1/\Av_{\deltav^*}(d) = (1+1/g)N s^2_d$ in formula (\ref{hfun2}) yields
$$
 h(\hat\alpha_d, g)
 =
 -\frac{N\hat\alpha^2_ds_d^2}{\verror}\frac 1{1+1/g} + \log(1+g)\,.
$$

Finally, as for the f-slab $\av_{0, \deltav^*}(d) =  \av_{\deltav^*}(d)$, $\Av_{0, \deltav^*}(d) = 1/(bNs^2_d)$ and $1/\Av_{\deltav^*}(d) = N s^2_d$, we have
$$
 \frac{\big(\av_{\deltav^*}(d)\big)^2}{\Av_{\deltav^*}(d)}
 - \frac{\big(\av_{0, \deltav^*}(d)\big)^2}{\Av_{0, \deltav^*}(d)}
 =
 (1-b) N\hat\alpha^2_d s_d^2
$$
and hence
$$
 h(\hat\alpha_d, b)
 = -(1-b)\frac{N\hat\alpha^2_ds_d^2}{\verror} - \log(b)\,.
$$

\subsection{Correlated Regressors}\label{app:corr}

We assume $s^2_j = 1$, $j = 1, 2$. To compute the posterior inclusion probability of $\xv_2$ when $\xv_1$ is included in the model we compare the conditional marginal likelihoods of the two models $\deltav = (1, 1)$ and $\deltav^* = (1, 0)$ by
\begin{eqnarray*}
 2\log\frac{p(\yv|\deltav^*,\verror)}{p(\yv|\deltav,\verror)}
 &=&
 -\frac 1{\verror}
 \left(\av_{0, \deltav^*}'\Av_{0, \deltav^*}^{-1}\av_{0, \deltav^*}
     - \av_{\deltav^*}'\Av_{\deltav^*}^{-1}\av_{\deltav^*}
     - \av_{0, \deltav}'\Av_{0, \deltav}^{-1}\av_{0, \deltav}
     + \av_{\deltav}'\Av_{\deltav}^{-1}\av_{\deltav}\right) \\
 & &
 + \log\frac{|\Av_{0, \deltav}||\Av_{\deltav^*}|}
            {|\Av_{0, \deltav^*}||\Av_{\deltav}|}\,.
\end{eqnarray*}
This simplifies as follows:
\begin{align*}
 \text{i-slab:}\quad
 2\log\frac{p(\yv|\deltav^*,\verror)}{p(\yv|\deltav,\verror)}
 & = -\frac 1{\verror}\left(\av_{\deltav}'\Av_{\deltav}^{-1}\av_{\deltav}
                     - \av_{\deltav^*}'\Av_{\deltav^*}^{-1}\av_{\deltav^*}\right)
                     + \log\frac{c|\Av_{\deltav^*}|}{|\Av_{\deltav}|}\\
 \text{g-slab:}\quad
 2\log\frac{p(\yv|\deltav^*,\verror)}{p(\yv|\deltav,\verror)}
 & = -\frac 1{\verror}\left(\av_{\deltav}'\Av_{\deltav}^{-1}\av_{\deltav}
                     - \av_{\deltav^*}'\Av_{\deltav^*}^{-1}\av_{\deltav^*}\right)
                     + \log(g+1)\\
 \text{f-slab:} \quad
 2\log\frac{p(\yv|\deltav^*,\verror)}{p(\yv|\deltav,\verror)}
 & = -\frac 1{\verror}\left(\av_{\deltav}'\Av_{\deltav}^{-1}\av_{\deltav}
                     - \av_{\deltav^*}'\Av_{\deltav^*}^{-1}\av_{\deltav^*}\right)
                     (1-b) - \log(b)\,.
\end{align*}

We give details for the g-slab. Note that using the notation introduced in Section \ref{sec:theo},
$$
 \Xv'\Xv = N
 \begin{pmatrix} 1 & r_{12} \\ r_{12} & 1\end{pmatrix}
 \qquad
 \text{and}
 \qquad
 \Xv'\yv_c = N s_y
 \begin{pmatrix} r_{y1} \\ r_{y2} \end{pmatrix}\,.
$$
$\deltav^*$ denotes the model with $\xv_1$ as the only regressor, hence we get (as for orthogonal regressors)
$$
 \av_{\deltav^*}'\Av_{\deltav^*}^{-1}\av_{\deltav^*}
 =
 \frac g{g+1}N r_{y1}^2s_y^2\,.
$$
As the corresponding term for model $\deltav$ is given as
$$
 \av_{\deltav}'\Av_{\deltav}^{-1}\av_{\deltav}
 =
 \frac g{g+1}\yv_c'\Xv(\Xv'\Xv)^{-1}\Xv'\yv_c
 =
 \frac g{g+1}\frac{Ns^2_y}{1-r_{12}^2}\left(r_{y1}^2 - 2r_{12}r_{y1}r_{y2}
 + r_{y2}^2\right)\,,
$$
we get
$$
 \av_{\deltav}'\Av_{\deltav}^{-1}\av_{\deltav}
 -
 \av_{\deltav^*}'\Av_{\deltav^*}^{-1}\av_{\deltav^*}
 =
 \frac g{g+1}\frac{Ns_y^2(r_{y2} - r_{y1}r_{12})^2}{1-r_{12}^2}\,,
$$
and finally, using $\hat\alpha_2 = \frac{s_y(r_{y2} - r_{12}r_{y1})}{(1 - r_{12}^2)}$, we obtain
$$
 2\log\frac{p(\yv|\deltav^*,\verror)}{p(\yv|\deltav,\verror)}
 = - \frac{N \hat\alpha_2^2}{\verror}(1 - r_{12}^2)\, \frac g{g+1}
 + \log(g+1)\,.
$$

Obviously for the f-slab we have
$$
 2\log\frac{p(\yv|\deltav^*,\verror)}{p(\yv|\deltav,\verror)}
 = -\frac{N\hat\alpha_2^2}{\verror}(1 - r_{12}^2)(1-b) - \log(b)\,.
$$
\end{appendix}

\bibliographystyle{apacite}
\bibliography{wallibib}

%


\end{document}